\documentclass[apj]{emulateapj}

\def\d{\delta}

\def\c{c}
\def\d{d}
\def\e{e}
\usepackage[dvipdfmx]{color}
\usepackage{bm}

\usepackage{aas_macros} 
\usepackage{float}
\usepackage{amsmath,amssymb}
\usepackage{slashbox}
\usepackage{url}
\slugcomment{Accepted 2013 October 21}
\shortauthors{Masuda et al.}
\shorttitle{Characterization of the KOI-94 system with TTV analysis
}
\begin{document}
\title{CHARACTERIZATION OF THE KOI-94 SYSTEM WITH TRANSIT TIMING VARIATION ANALYSIS: 
IMPLICATION FOR THE PLANET-PLANET ECLIPSE}
\author{
Kento \textsc{Masuda},\altaffilmark{1}
Teruyuki \textsc{Hirano},\altaffilmark{2}
Atsushi \textsc{Taruya},\altaffilmark{1,3,4}
Makiko\ \textsc{Nagasawa},\altaffilmark{5}
and 
Yasushi \textsc{Suto}\altaffilmark{1,3,6}
} 
\altaffiltext{1}{
Department of Physics, The University of Tokyo, Tokyo 113-0033, Japan
}

\altaffiltext{2}{
Department of Earth and Planetary Sciences, Tokyo Institute of Technology,
2-12-1 Ookayama, Meguro-ku, Tokyo 152-8551, Japan
}

\altaffiltext{3}{
Research Center for the Early Universe, Graduate School of Science, 
The University of Tokyo, Bunkyo-ku, Tokyo 113-0033, Japan
}

\altaffiltext{4}{
Yukawa Institute for Theoretical Physics, Kyoto University, Kyoto 606-8502, Japan}

\altaffiltext{5}{
Interactive Research Center of Science, Tokyo Institute of Technology,
2-12-1 Ookayama, Meguro-ku, Tokyo 152-8551, Japan
}

\altaffiltext{6}{
Department of Astrophysical Sciences, Princeton
University, Princeton, NJ 08544, USA}

\email{masuda@utap.phys.s.u-tokyo.ac.jp}
\begin{abstract}
The KOI-94 system is a closely-packed, multi-transiting planetary system 
discovered by the {\it Kepler} space telescope.
It is known as the first system that exhibited a rare event called a ``planet-planet eclipse (PPE),"  
in which two planets partially overlap with each other in their double-transit phase.
In this paper, we constrain the parameters of the KOI-94 system with an analysis of the transit timing variations (TTVs).
Such constraints are independent of the radial velocity (RV) analysis recently performed by Weiss and coworkers,
and valuable in examining the reliability of the parameter estimate using TTVs.
We numerically fit the observed TTVs of KOI-94c, KOI-94d, and KOI-94e for their masses, eccentricities, and longitudes of periastrons,
and obtain the best-fit parameters including
$m_{\rm c} = 9.4_{-2.1}^{+2.4} M_{\oplus}$, 
$m_{\rm d} = 52.1_{-7.1}^{+6.9} M_{\oplus}$,
$m_{\rm e} = 13.0_{-2.1}^{+2.5} M_{\oplus}$, and $e \lesssim 0.1$ for all the three planets.
While these values are mostly in agreement with the RV result,
the mass of KOI-94d estimated from the TTV is significantly smaller than the RV value
$m_{\rm d} = 106 \pm 11 M_{\oplus}$.
In addition, we find that the TTV of the outermost planet KOI-94e is not well reproduced in the current modeling.
We also present analytic modeling of the PPE and derive a simple formula to reconstruct 
the mutual inclination of the two planets from
the observed height, central time, and duration of the brightening caused by the PPE.
Based on this model, the implication of the results of TTV analysis for the time of the next PPE 
is discussed.
\end{abstract}
\keywords{
planets and satellites: individual (KOI-94, KIC 6462863, Kepler-89) -- 
techniques: photometric
}
\section{Introduction} \label{sec:intro}
\begin{figure}
	\centering
	\includegraphics[width=7.5cm,clip]{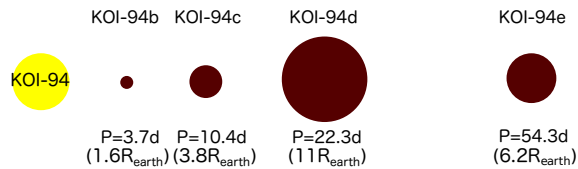}
	\caption{Schematic illustration of the KOI-94 system. Planetary radii are calculated from the planet-to-star radius ratio,
	stellar density obtained from the transit light curves, and spectroscopic stellar mass $M_{\star} = 1.25 M_{\odot}$
	\citep{2012ApJ...759L..36H}.
	}
	\label{koi94}
\end{figure}

The Kepler Object of Interest (KOI) 94 system is a multi-transiting planetary system discovered by the {\it Kepler} space telescope 
\citep{2011ApJ...736...19B, 2013ApJS..204...24B},
consisting of four transiting planets with periods of about
3.7 (KOI-94b), 10 (KOI-94c), 22 (KOI-94d), and 54 (KOI-94e) days (Figure \ref{koi94}).
For the largest planet KOI-94d, \citet{2012ApJ...759L..36H} observed the Rossiter-McLaughlin effect
\citep[][]{1924ApJ....60...15R, 1924ApJ....60...22M, 2000A&A...359L..13Q, 2005ApJ...622.1118O, 2005ApJ...631.1215W, 2011ApJ...742...69H}
for the first time in a multi-transiting system.
They found $\lambda = -6_{-11}^{+13}\,\mathrm{deg}$, showing that the orbital axis of this planet is
aligned with the stellar spin axis 
(this result was later confirmed by \citet{2013ApJ...771...11A}, 
who obtained $\lambda = -11 \pm 11\,\mathrm{deg}$).
Furthermore, the KOI-94 system is the first and only system in which a rare mutual event called 
a ``planet-planet eclipse" (hereafter PPE) was identified; in this event, two planets transit simultaneously 
and partially overlap with each other on the stellar disk as seen from our line of sight.
By analyzing the light curve of the PPE caused by KOI-94d and KOI-94e,
\citet{2012ApJ...759L..36H} concluded that the orbital planes of these two planets are also well aligned
within $2$ degrees.
In this system, therefore, the stellar spin axis and the orbital axes of the two planets are all aligned.
If their close-in orbits are due to planetary migration \citep[e.g.][]{2011exop.book..347L}, this result suggests that 
they have experienced a quiescent disk migration \citep{1980ApJ...241..425G} rather than processes that include
gravitational perturbations either by planets \citep[e.g.][]{2008ApJ...678..498N, 2011ApJ...735..109W, 2011Natur.473..187N} or stars \citep[e.g.][]{2003ApJ...589..605W}.
Other processes that tilt the stellar spin axis relative to orbital axes of planets 
\citep[e.g.][]{2010MNRAS.401.1505B, 2011MNRAS.412.2790L, 2012ApJ...758L...6R, 2012Natur.491..418B} are also excluded,
provided that the orbital planes of the multiple transiting planets trace the original protoplanetary disk
from which they formed.
For these reasons, the KOI-94 system is an important test bed that provides a clue to understand the formation 
process of closely-packed multi-transiting planetary systems, and hence deserves to be characterized in detail.

Recently, \citet{2013ApJ...768...14W} measured the radial velocities (RVs) of KOI-94 from the W. M. Keck Observatory,
and estimated the masses and eccentricities of the planets by a simultaneous fit to the observed RVs and the {\it Kepler} light curve. 
They showed that the masses of all the planets fall into the planetary regime,
and especially obtained a fairly well constraint on the mass of KOI-94d 
($m_{\rm d} = 106 \pm 11 M_{\oplus}$).
However, the masses of KOI-94c ($m_{\rm c} = 15.6_{-15.6}^{+5.7} M_{\oplus}$) and
KOI-94e ($m_{\rm e} = 35_{-28}^{+18} M_{\oplus}$) are weakly constrained
because of the marginal detections of their RV signals.
In addition, the best-fit eccentricity of KOI-94c ($e_{\rm c} = 0.43 \pm 0.23$) is suspiciously large in light of the 
long-term stability of the system, as pointed out in their paper.
Hence additional RV observations are definitely important, but transit timing variations
\citep[TTVs,][]{2005Sci...307.1288H, 2005MNRAS.359..567A}
can also be used to improve these estimates
in such a multi-transiting system like KOI-94
\citep[e.g. TTV analysis in the Kepler-11 system by][]{2011Natur.470...53L}.
Moreover, in the KOI-94 system the orbital parameters are exceptionally well constrained by the observations of the 
Rossiter-McLaughlin effect and the PPE; this makes the KOI-94 system an ideal case to evaluate the reliability of
the parameter estimates by TTVs in comparison to RVs.

Apart from such characterization of the KOI-94 system, the PPE itself is a unique phenomenon that is worth studying in a more general context.
If this event is observed in the future transit observations, it can be used to precisely constrain the relative angular momentum
of the planets, which is closely related to their orbital evolution processes.
In fact, this phenomenon had been theoretically predicted before by \citet{2010arXiv1006.3727R}
\citep[see also][]{2009A&A...494..391R} as an ``overlapping double transit," 
and they emphasized its role in constraining the relative nodal angle of the planets.
However, neither the analytic formulation 
that clarifies the physical picture of this phenomenon, nor the discussion about how gravitational interactions 
among the planets affect the PPE, has been presented so far.

In this paper, we investigate the constraints on masses and eccentricities of KOI-94c, KOI-94d, and KOI-94e
based on the direct numerical analysis of their TTV signals.
We also construct an analytic model of the PPE, and discuss how the gravitational interaction affects the occurrence of the next PPE
based on the model and the result of TTV analysis.

The plan of this paper is as follows. 
First, we perform an intensive TTV analysis in Section \ref{sec:photometry}.
We discuss the constraints on
transit parameters based on the phase-folded transit light curves, and those on the mass, eccentricity, and
longitude of periastron based on the numerical fit to the observed TTV signals.
Then in Section \ref{sec:PPE_form}, we present an analytic description of the PPE which elucidates 
how the height, duration, and central time of the brightening caused by the overlap are related to the orbital parameters.
Based on this formulation, we provide a general procedure for constraining the orbits of the overlapping planets in Section \ref{sec:PPE_app}.
Here we also discuss a simple prediction of the next PPE on the basis of a two-body problem.
Finally, based on the analytic model of the PPE and the result of TTV analysis,  
we show in Section \ref{sec:ppnext} how the gravitational interaction among the planets affects the occurrence of the next PPE 
in the KOI-94 system, referring to the difference from the two-body prediction.
Section \ref{sec:summary} summarizes the paper.
The results on the properties of the KOI-94 system are all in Section \ref{sec:photometry},
and so the readers who are only interested in the TTV analysis can skip Sections \ref{sec:PPE_form} to \ref{sec:ppnext},
where we mainly discuss the PPE.
\section{Analysis of the Photometric Light Curves} \label{sec:photometry}
In this section, we report the analysis of photometric light curves of KOI-94 taken by {\it Kepler}.
We determine the orbital phases, scaled semi-major axes, scaled planetary radii, and inclinations 
of KOI-94c, KOI-94d, and KOI-94e from the phase-folded transit light curves,
and estimate their masses, eccentricities, and longitudes of periastrons from their TTV signals (see Table \ref{ppar_def}). 
In the following analysis, we neglect the smallest and innermost planet KOI-94b, 
which does not affect the TTV signals of the other three, as we will see in Section \ref{sec:Nbody_RV}.
\begin{table*}
	\begin{center}
	\caption{Definitions of the parameters derived in this paper.}
	\label{ppar_def}
	\begin{tabular}{cc}
	\tableline \tableline
	 Parameter			&	Definition	\\
	\tableline
	\multicolumn{2}{c}{\it Parameters derived from transit light curves}\rule[-1.2mm]{0mm}{4mm}\\
	\tableline
	$t_0$				& 	Time of a transit center (BJD - 2454833)\\
	$P$					& 	Orbital period\\
	$R_p / R_{\star}$ 		&	Planet-to-star radius ratio\\
	$a / R_{\star}$ 			&	Scaled semi-major axis\\          
	$b $ 					&    	Impact parameter of the transit ($= a \cos i / R_{\star}$, $i$: orbital inclination)\\
	$u_1$, $u_2$ 			& 	Coefficients for the quadratic limb-darkening law\\
	\tableline 
	\multicolumn{2}{c}{\it Parameters derived from the PPE\,\tablenotemark{a}}\rule[-1.2mm]{0mm}{4mm}
	\\
	\tableline
	$\Omega$			& 	Longitude of the ascending node\\
	\tableline
	\multicolumn{2}{c}{\it Parameters derived from TTVs}\rule[-1.2mm]{0mm}{4mm}\\
	\tableline
	$m$					&	Planetary mass\\
	$e$					&	Orbital eccentricity\\
	$\varpi = \omega + \Omega$				&	Longitude of the periastron\\
	\tableline
	\end{tabular}
	\tablenotetext{1}{PPE can only constrain the difference of the nodal angles of KOI-94d and KOI-94e. We have no information
	on $\Omega$ of KOI-94c.}
	\end{center}
\end{table*}
\begin{table*}
	\begin{center}
	\caption{Parameter values estimated by other authors.}
	\label{ppar_prior}
	\begin{tabular}{cccc}
	\tableline \tableline
	 Parameter			&	KOI-94c					&	KOI-94d					&	KOI-94e\\				
	\tableline
	\multicolumn{4}{c}{\it Transit parameters determined by the Kepler team\tablenotemark{a}}\rule[-1.2mm]{0mm}{4mm}\\
	\tableline
	$t_0$ (BJD - 2454833) & $138.00718 \pm 0.00093$ 		& $132.74047 \pm 0.00019$ 		& $161.23998 \pm 0.00079$\\
	$P$ (days)			& $10.423707 \pm 0.000026$ 	& $22.343001 \pm 0.000011$		& $54.31993 \pm 0.00012$\\
	$a / R_{\star}$ 		& $15.70 \pm 0.37$	 			& $26.10 \pm 0.62$ 			& $47.2 \pm 1.1$\\
	$R_p / R_{\star}$ 	& $0.02544 \pm 0.00012$ 		& $0.06856 \pm 0.00012$ 		& $0.04058 \pm 0.00013$\\
	$b$ 				& $0.019 \pm 0.048$ 			& $0.305 \pm 0.014$ 			& $0.387 \pm 0.014$\\
	\tableline 
	\multicolumn{4}{c}{\it  Parameters determined by \citet{2012ApJ...759L..36H}}\rule[-1.2mm]{0mm}{4mm}\\
	\tableline
	$\Omega$ (deg)\tablenotemark{b}	& $-6_{-11}^{+13}$		& $-$ 			& $-5_{-11}^{+13}$	\\
	$u_1$			& \multicolumn{3}{c}{$0.10 \pm 0.06$}\\
	$u_2$			& \multicolumn{3}{c}{$0.61 \pm 0.08$}\\
	\tableline
	\multicolumn{4}{c}{\it  Parameters determined by \citet{2013ApJ...768...14W}}\rule[-1.2mm]{0mm}{4mm}\\
	\tableline
	$m$ ($M_{\oplus}$)	& $15.6_{-15.6}^{+5.7}$			& $106 \pm 11$				& $35_{-28}^{+18}$\\	
	$e$				& $0.43 \pm 0.23$				& $0.022 \pm 0.038$			& $0.019 \pm  0.23$\\
	\tableline
	\end{tabular}
	\tablenotetext{1}{Data from MAST archive \url{http://archive.stsci.edu/kepler/}}
	\tablenotetext{2}{In this paper, we define the reference direction so that the spin-orbit angle $\lambda$ 
	measured by the Rossiter-McLaughlin effect be equal to $\Omega$ for the orbital inclination in the range $[0, \pi/2]$ \citep{2009ApJ...696.1230F}.
	With this choice, the reference line points to the ascending node of a virtual circular orbit whose angular momentum is parallel to the 
	stellar spin vector.}
	\end{center}
\end{table*}
\subsection{Transit times and transit parameters}
\subsubsection{Data processing} \label{sec:dataprocess}
We analyze the short-cadence ($\sim 1$ min) PDCSAP 
(Pre-search Data Conditioned Simple Aperture Photometry) fluxes \citep[e.g.][]{2012PASP..124..963K}
from Quarters $4$, $5$, $8$, $9$, $12$, and $13$.
We do not include the data from Quarter $1$, for which only the long-cadence data is available.
Since these light curves exhibit the long-term trends that affect the baseline of the transit,
we remove those trends in the following manner.
First, data points within $\pm 1$ day of every transit caused by KOI-94c, KOI-94d, or KOI-94e are extracted 
and each set of the data is fitted with a fifth-order polynomial, masking out the points during the transit.
Then we calculate the standard deviation of each fit, 
remove outliers exceeding $5\sigma$, and fit the data again with the fifth-order polynomial.
This process is iterated until all the $5\sigma$ outliers are removed.
Finally, all the data points in each chunk (including those during the transit) are divided by the best-fit polynomial
to yield a detrended and normalized transit light curve.
In our analysis of the TTV, we exclude the transits whose ingress or egress is not completely observed
due to the data cadence of {\it Kepler}.
We also exclude the ``double-transit" events, during which two planets transit the stellar disk at the same time.
As an exception, the double transit of KOI-94d and KOI-94e around BJD = $2454211.5$
(in which a PPE was observed) is included in our analysis;
in this case the ingresses and egresses of both transits are clearly seen because of their close mid-transit times.
The above criteria leave us with $44$, $21$, and $8$ transits for KOI-94c, KOI-94d, and KOI-94e, respectively.

\subsubsection{Transit parameters}
Before analyzing the TTV signals, we revise the transit parameters of 
KOI-94c, KOI-94d, and KOI-94e obtained by the Kepler team (Table \ref{ppar_prior})
so that they are consistent with the light curves obtained in Section \ref{sec:dataprocess}.
Here we first use the parameters publicized by the Kepler team to phase-fold the observed transit light curves,
and then refit those phase curves to obtain the revised transit parameters.

In the first step, we fit each of the detrended light curve centered at the transit (for $\sim 1.7$ times its duration)
to obtain the times of transit centers $t_c$, using a Markov chain Monte Carlo (MCMC) algorithm.
Here we use a light curve model by \citet{2009ApJ...690....1O}, 
and fix $a/R_{\star}$, $R_p/R_{\star}$, and $b$ to the values obtained by the Kepler team,
assuming $e = 0$.
We model the limb-darkening using a quadratic law in Eq.(\ref{qlimb}) 
and adopt the limb-darkening coefficients $u_1$ and $u_2$ obtained by \citet{2012ApJ...759L..36H} (all these parameters are summarized in Table \ref{ppar_prior}).
Since the detrend procedure in Section \ref{sec:dataprocess} can remove only the out-of-transit outliers, 
we also exclude in-transit $5\sigma$ outliers of this fit, if any, and fit the light curve again.
Using the series of $t_c$ obtained in this way, we construct the phase-folded transit light curve for each planet.

\begin{figure}
	\begin{center}
	\includegraphics[width=7.3cm,clip]{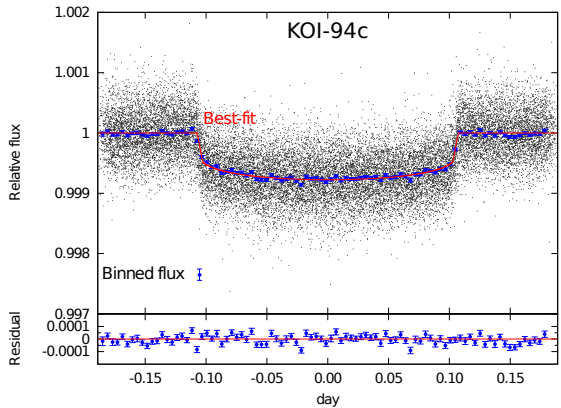}	
	\caption{Phase-folded transit light curve of KOI-94c.
	The best-fit model is shown with the red solid line.
	Small gray dots are all the short-cadence data points.
	Blue points are fluxes binned to $0.1\,\mathrm{hr}$ bins and their error bars are calculated by
	$1.4826 \times {\rm median\ absolute\ deviation}$.}
	\label{94c_phase}
	\end{center}
\end{figure}

\begin{figure}
	\begin{center}
	\includegraphics[width=7.3cm,clip]{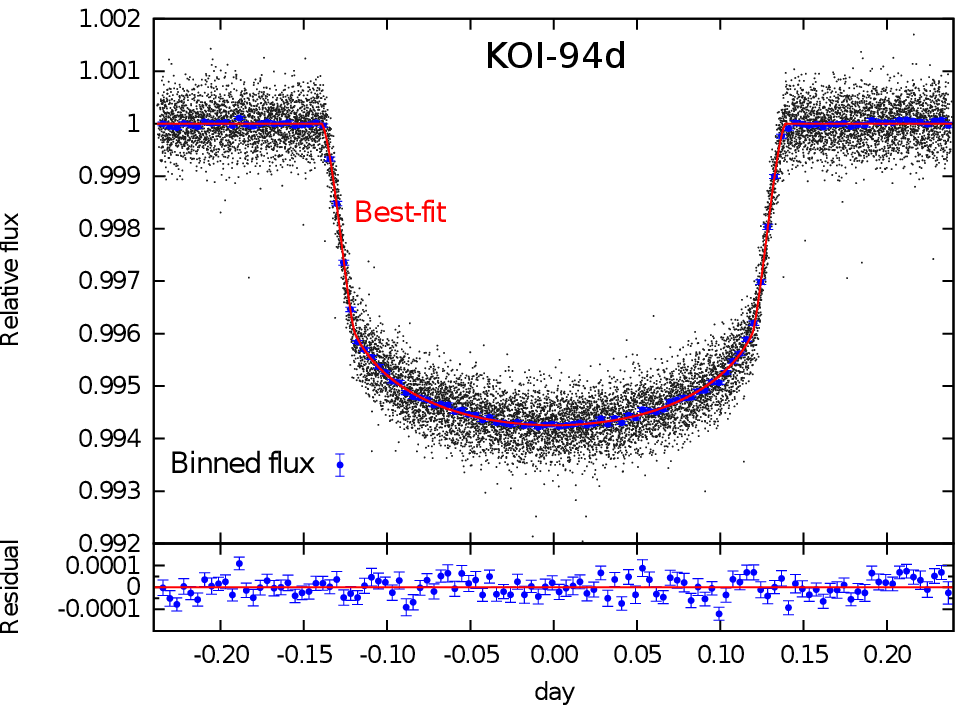}		
	\caption{Phase-folded transit light curve of KOI-94d (same as Figure \ref{94c_phase}).}
	\label{94d_phase}
	\end{center}
\end{figure}

\begin{figure}
	\begin{center}
	\includegraphics[width=7.35cm,clip]{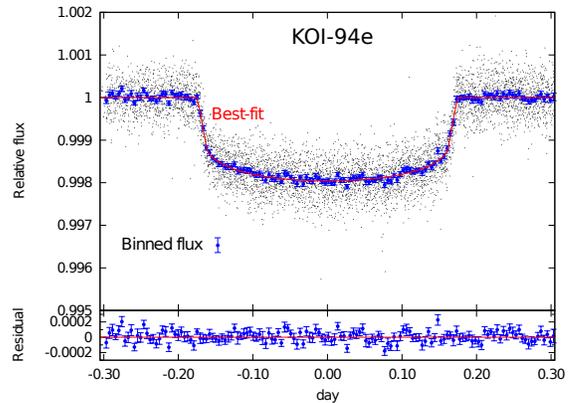}		
	\caption{Phase-folded transit light curve of KOI-94e (same as Figure \ref{94c_phase}).}	
	\label{94e_phase}
	\end{center}
\end{figure}

As the second step, we fit the resulting phase-folded transit light curve for $a/R_{\star}$, $R_p/R_{\star}$, $b$, $u_1$, and $u_2$ 
using the same light curve model as above.
In this way, we obtain the revised values of the set of parameters shown in Table \ref{ppar_fit} and the corresponding best-fit 
light curves (Figures \ref{94c_phase} to \ref{94e_phase}).\footnote{
We also repeated the same analysis taking account of the quarter-to-quarter flux contaminations publicized by the Kepler team and available at MAST archive.
As expected, we obtained larger $R_p$ by a fraction of $\sim c/2$, where $c$ is the fractional contamination \citep[e.g.][]{2012ApJ...750..114F}, 
but the other parameters were consistent within $2.1 \sigma$ except for $a_{\rm c}$ and $a_{\rm e}$.
These two parameters were different from those in Table \ref{ppar_fit} by $\sim 1\%$, corresponding to a slight change in the value of $a_{\rm d}$.
In this paper, we do not apply this correction because the smaller values of $R_p$ lead to more conservative 
estimates for the PPE occurrence.}
In this fit, all the parameters converge well in the case of KOI-94d.
In contrast, $a/R_{\star}$ and $b$ of KOI-94c and KOI-94e do not converge well moving back and forth between several local minima
in a strongly correlated fashion, because they show smaller transit depths and the ingresses/egresses of their transits are less clear.
For this reason, we impose an additional constraint 
that all the planets share the same host star:
we convert the well-constrained $a/R_{\star}$ for KOI-94d into stellar density $\rho_{\star}$ 
via $\rho_{\star} \approx (3 \pi /GP^2) (a/R_{\star})^3$ \citep{2003ApJ...585.1038S},
and calculate the corresponding values and uncertainties of $a/R_{\star}$ for KOI-94c and KOI-94e.
The phase curves of KOI-94c and KOI-94e are fitted with prior constraints centered on these values and
with Gaussian widths of their uncertainties.
With this prescription, all the parameters of KOI-94c and KOI-94e converge well.
Therefore, it does not make sense here to discuss the consistency of $\rho_{\star}$ calculated from the transit parameters 
to check the possible false positives.
It is important to note, however, that the limb-darkening coefficients for each planet obtained individually
are consistent within their $1\sigma$ error bars
(those obtained by \citet{2012ApJ...759L..36H} are different from our values because they fixed smaller $R_p/R_{\star}$
for KOI-94d; see Table \ref{ppar_prior}).
This supports the notion that these three planets are indeed revolving around the same host star.
\begin{table*}
	\begin{center}
	\caption{Revised transit parameters obtained from phase-folded light curves.}
	\label{ppar_fit}
	\begin{tabular}{cccc}
	\tableline \tableline
	 					&		KOI-94c						&	KOI-94d							&	KOI-94e							\\
	\tableline
	$a / R_{\star}$ 			& $15.7798_{-0.0016}^{+0.0016}$			& $26.24_{-0.19}^{+0.20}$ 				& $47.4400_{-0.0093}^{+0.0094}$	\\
	$R_p / R_{\star}$ 		& $0.025673_{-0.000075}^{+0.000074}$	& $0.07029_{-0.00015}^{+0.00014}$ 		& $0.04137_{-0.00013}^{+0.00012}$		\\
	$b$ 					& $0.021_{-0.015}^{+0.020}$				& $0.299_{-0.025}^{+0.022}$ 				& $0.3840_{-0.0055}^{+0.0048}$ 			\\
	$u_1$ 				& $0.40 \pm 0.06$ 						& $0.40 \pm 0.02$ 						& $0.36 \pm 0.07$ 						\\
	$u_2$ 				& $0.10_{-0.09}^{+0.08}$ 				& $0.14 \pm 0.03$ 						& $0.19_{-0.10}^{+0.11}$ 				\\
	$\chi^2/\mathrm{d.o.f}$	& $25843/23611$						& $15473/13766$ 						& $6822/5934$						\\
	\tableline 
	\end{tabular}
	\tablecomments{The quoted error bars denote $1\sigma$ confidence intervals obtained from the posterior distributions.
	The values of $a/R_{\star}$ and $b$ for KOI-94c and KOI-94e are determined with the prior information about the stellar density
	based on the result for KOI-94d (see the main text).}
	\end{center}
\end{table*}
\subsubsection{TTV signals}
\begin{figure}
	\begin{center}
	\includegraphics[width=7.5cm,clip]{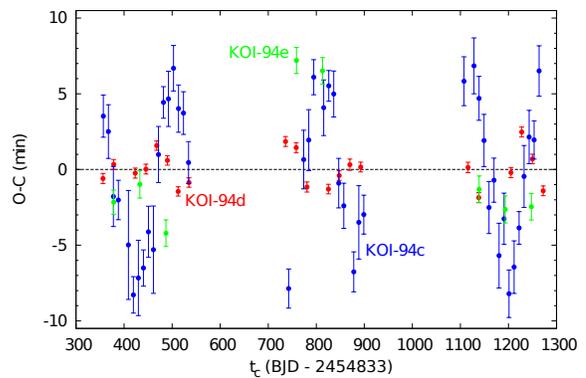}
	\caption{Observed TTV signals of  KOI-94c (blue), KOI-94d (red), and KOI-94e (green).}
	\label{ttvsignal}
	\end{center}
\end{figure}

Fixing $a/R_{\star}$, $R_p/R_{\star}$, $b$, $u_1$, and $u_2$ at the values in Table \ref{ppar_fit},
we refit the transit light curves of each planet to find the values of $t_c$ given in Tables \ref{94.02tc} to \ref{94.03tc} 
in Appendix \ref{sec:tc},
with the values of reduced $\chi^2$ and $1\sigma$ upper/lower limits obtained from the posterior. 
The column labeled as $O-C$ tabulates the residuals of a linear fit to $t_c$ versus transit number,
in which linear ephemerides in Table \ref{ep} are extracted.
The values of reduced $\chi^2$ of the linear fits in this table indicate the significant deviations of the transit times 
from the linear ephemerides (i.e. TTVs) for all the three planets, as shown in Figure \ref{ttvsignal}.
Note that the TTV of KOI-94c shows the modulation with the period of 
$(1/ P_{\rm c} - 2 / P_{\rm d})^{-1} \simeq 155\,{\rm days}$, which clearly comes from
near 2:1 resonance of KOI-94c and KOI-94d as pointed out by \citet{2013arXiv1308.3751X} (see also Appendix \ref{sec:lithwick}).
\begin{table*}
	\begin{center}
	\caption{Linear ephemerides of KOI-94{\rm \c}, KOI-94{\rm \d}, and KOI-94{\rm \e}.}
	\label{ep}
	\begin{tabular}{cccc}
	\tableline \tableline
	Parameter					&	KOI-94c						& KOI-94d					& KOI-94e	\\
	\tableline
	$t_0$ (${\rm BJD} - 2454833$)		&	$138.00826 \pm 0.00038$		& $132.74103 \pm 0.00012$		& $161.23888 \pm 0.00046$	\\
	$P$ (days)					&	$10.4236888 \pm 0.0000053$	& $22.3429698 \pm 0.0000036$	& $54.319849 \pm 0.000035$	\\
	$\chi^2/\mathrm{d.o.f}$		&	$10.4$						& $13.7$						& $29.2$\\
	\tableline 
	\end{tabular}
	\end{center}
\end{table*}
\subsection{Numerical analysis of the TTV signals using RV mass of KOI-94d} \label{sec:Nbody_RV}
We numerically analyze the TTV signals in Figure \ref{ttvsignal} to constrain the masses, eccentricities,
and longitudes of periastrons of KOI-94c, KOI-94d, and KOI-94e (nine parameters in total).
In this section, we fix the mass of KOI-94d at $m_{\rm d} = 106 M_{\oplus}$, the best-fit RV value obtained by \citet{2013ApJ...768...14W}.
Since this value is only marginally consistent with $m_{\rm d} = 73 \pm 25 M_{\oplus}$ obtained by \citet{2012ApJ...759L..36H} 
based on the out-of-transit RVs taken by the Subaru telescope, 
we also investigate the case of $m_{\rm d} = 73 M_{\oplus}$ here.

\subsubsection{Calculation of the simulated TTV} 
Orbits of the planets are integrated using the fourth-order Hermite scheme with the shared time step \citep{2004PASJ...56..861K}.
The time step (typically $\sim 0.025$ days) is chosen so that the fractional energy change due to the integrator
during $\sim 1000$ days integration should always be smaller than $10^{-9}$.
All the simulations presented in this section integrate planetary orbits beginning at the same epoch 
$T_0 (\mathrm{BJD}) = 2455189.1703$ (the first transit time of KOI-94d), 
until $\mathrm{BJD} = 2456133.0$ (approximately the last transit time of KOI-94d we analyzed).

For each planet, the initial value of the orbital inclination $i$ is fixed at the value obtained from $a/R_{\star}$ and $b$ (see Table \ref{ppar_fit}), 
assuming that $e = 0$.
Here, all the $i$ values are chosen in the range $[0, \pi/2]$.\footnote{ 
As we will see in Section \ref{sec:PPE_app}, the observed PPE light curve requires that if we (arbitrarily) choose $i_{\rm d}$ in $[0, \pi/2]$, $i_{\rm e}$ is
also in this range. There is no justification to choose $i_{\rm c}$ also in this range, but this choice does not affect the result significantly
because the value of $i_{\rm c}$ is very close to $\pi/2$, as suggested by the small value of $b_{\rm c}$ (c.f. Table \ref{ppar_fit})}
The values of $\Omega_{\rm d}$ and $\Omega_{\rm e}$ are fixed at those in Table \ref{ppar_prior}.
Since we have no information on the nodal angle of KOI-94c, we assume $\Omega_{\rm c} = 0$ deg.
Initial semi-major axes are calculated via Kepler's third law
with $M_{\star} = 1.25 M_{\odot}$ \citep{2012ApJ...759L..36H}, orbital periods in Table \ref{ep},
and planetary masses adopted in each simulation.
The phases of the planets are determined from the transit ephemerides:
for each planet, we convert the transit time closest to $T_0$ into the sum of the argument of periastron and mean anomaly $\omega + f$,
taking account of the non-zero eccentricity if any,
and then move it backward in time to $T_0$, assuming a Keplerian orbit.

The mid-transit times of each planet are determined by minimizing the sky-plane distance $D$ between the star and the planet,
where the roots of the time derivative of $D$ are found by the Newton-Raphson method \citep{2010arXiv1006.3834F}.
Then these transit times are fitted with a straight line
and thereby the TTVs ($=$ residuals of the linear fit), as well as the linear ephemeris ($P$ and $t_0$), 
are extracted.
We compute the chi squares of the simulated TTVs obtained in this way as
\begin{equation}
	\chi^2_{j} = \sum_{\substack{i:{\rm observed}\\{\rm transits}}}
			\left[ \frac{{\rm TTV}^{(j)}_{\rm sim}(i) -  {\rm TTV}^{(j)}_{\rm obs}(i)}{\sigma^{(j)}_{\rm obs}(i)} \right]^2,
	\quad (j = {\rm c}, {\rm d}, {\rm e})
\end{equation} 
where ${\rm TTV}^{(j)}_{\rm sim}(i)$ and ${\rm TTV}^{(j)}_{\rm obs}(i)$ are the $i$-th values of simulated and observed TTVs of
planet $j$, respectively, and $\sigma^{(j)}_{\rm obs}(i)$ is the observational uncertainty of the $i$-th transit time of planet $j$. 

Note that we do not fit the transit times directly but only the {\it deviations} from the periodicity in our analysis,
assuming that they provide sufficient information on the gravitational interaction among the planets.
Indeed, although the initial values of semi-major axes are chosen to match the observed periods,
periods derived from the simulations are different typically by $\sim 0.01$ days.
This is because strong gravitational interaction among massive, closely-packed planets in this system
causes the oscillations of their semi-major axes with amplitudes dependent on the 
parameters of the planets adopted in each run.
We will show that this simplified method still yields reasonable results in the last part of Section \ref{sec:ttv_discussion}.

\subsubsection{Estimates for the TTV amplitudes}
Before directly fitting the observed TTV signals, we evaluate the contribution from each planet
to the TTVs of KOI-94c, KOI-94d, and KOI-94e.
We divide the four planets into six pairs and integrate circular orbits for each pair
using the best-fit masses by \citet{2013ApJ...768...14W} listed in Table \ref{ppar_prior}.
Semi-amplitudes of the resulting TTVs of KOI-94c, KOI-94d, and KOI-94e are shown in Table \ref{ttv_pair}.
\begin{table*}
	\begin{center}
	\caption{Semi-amplitude of the simulated TTV (in units of min) for each planet pair.}
	\label{ttv_pair}
	\begin{tabular}{cccccc}
	\tableline \tableline
	\backslashbox{TTV}{pair}	&	KOI-94b	  &	KOI-94c	 &	   KOI-94d		&	KOI-94e		&  Major parameters for TTV\\
	\tableline
	KOI-94c 		&  $\lesssim 0.05$		&   -   		&    $11$			& $\lesssim 0.05$	& $m_{\rm d}$, $\bm{e}_{\rm c}$, $\bm{e}_{\rm d}$\\
	KOI-94d 		&  $\lesssim 0.05$ 		&   $0.47$		&    -				& $2.1$			& $m_{\rm c}$, $m_{\rm e}$, $\bm{e}_{\rm c}$, $\bm{e}_{\rm d}$, $\bm{e}_{\rm e}$\\
	KOI-94e 		&  $\lesssim 0.05$	 	&   $0.15$		&    $0.83$ 		&	-      			& $m_{\rm d}$, $\bm{e}_{\rm d}$, $\bm{e}_{\rm e}$, ($m_{\rm c}$, $\bm{e}_{\rm c}$)\\
	\tableline 
	\end{tabular}
	\end{center}
\end{table*}
Considering the uncertainties of transit times listed in Tables \ref{94.02tc} to \ref{94.03tc} (typically $1.4\,\mathrm{min}$, $0.3\,\mathrm{min}$, and 
$0.9\,\mathrm{min}$ for KOI-94c, KOI-94d, and KOI-94e, respectively),
this result indicates that KOI-94b has negligible contribution for the TTVs of the other three.
In the following analysis, therefore, we integrate the orbits of the other three planets (KOI-94c, KOI-94d, and KOI-94e) only.
We also find that the TTVs of KOI-94c and KOI-94e are mainly determined by the perturbation from the neighboring planet KOI-94d,
while that of KOI-94d depends on both of its neighbors.
Such dependence is naturally understood from the architecture of this system (see Figure \ref{koi94}).
Consequently, each planet's TTV mainly depends on the parameters listed in the rightmost column of Table \ref{ttv_pair},
where we define $\bm{e}_j = (e_j \cos \varpi_j, e_j \sin \varpi_j)$ ($j = {\rm c}, {\rm d}, {\rm e}$).
Note that the TTV of each planet is insensitive to its own mass. This is why $m_j$ ($j = {\rm c, d, e}$) is
not included in the row for planet $j$.

\subsubsection{Results} \label{sec:ttv_rv_result}
\begin{figure}
	\begin{center}
	\includegraphics[width=8.5cm,clip]{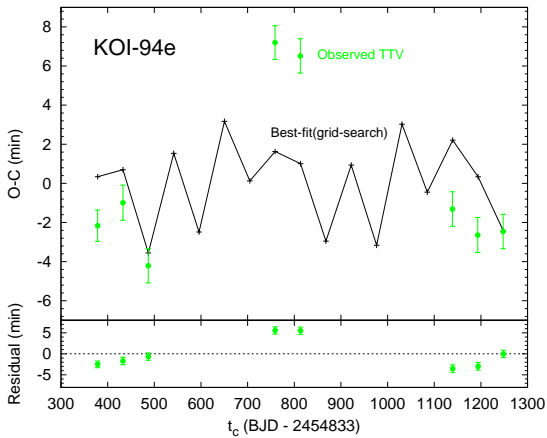}
	\caption{Best-fit simulated TTV of KOI-94e obtained by the grid-search for $m_{\rm d} = 106 M_{\oplus}$ 
	(black crosses connected with the solid line) with observed data (green points).
	The best-fit corresponds to $\bm{e}_{\rm d} = (0.02, 0.02)$ and $\bm{e}_{\rm e} = (0.04, 0.02)$.}
	\label{ttv03_heavy}
	\end{center}
\end{figure}

\noindent{\it TTV of KOI-94e}.\ ---
Since the TTV of KOI-94e is mainly determined by $\bm{e}_{\rm d}$ and $\bm{e}_{\rm e}$ (and $m_{\rm d}$, of course, which we fix at the RV value),
we fit it first so as to constrain these parameters.
We calculate $\chi^2_{\rm e}$ for 
$|e_{\rm d} \cos \varpi_{\rm d}|$, $|e_{\rm d} \sin \varpi_{\rm d}| \leq 0.06$ and
$|e_{\rm e} \cos \varpi_{\rm e}|$, $|e_{\rm e} \sin \varpi_{\rm e}| \leq 0.25$
(which well cover the $1\sigma$ regions for these parameters obtained from the RVs) at the grid-spacing of 0.01,
fixing $e_{\rm c} = 0$ and planetary masses at the best-fit values from the RVs.
However, we cannot fit the observed TTV well in both $m_{\rm d} = 106 M_{\oplus}$ and $73 M_{\oplus}$ cases.
The best-fit for the former case, which gives $\chi^2_{\rm e} = 122$ for $4$ degrees of freedom, is shown in Figure \ref{ttv03_heavy}.
For this reason, in addition to the fact that we have only eight transits observed for KOI-94e,
we decide not to use the TTV of this planet to constrain the system parameters, but fit only the TTVs of KOI-94c and KOI-94d.
The large discrepancy in the amplitudes of simulated and observed TTVs may suggest
another source of perturbation which is not included in our model, such as a non-transiting planet or other minor bodies.

\noindent{\it Grid-search for an initial parameter set}.\ ---
Then we perform the grid-search fit to the TTVs of KOI-94c and KOI-94d to find an appropriate initial parameter set
for the following MCMC analysis.
Based on the estimates given in Table \ref{ttv_pair}, we fit these TTVs separately as follows.
We first fit the TTV of KOI-94c varying $\bm{e}_{\rm c}$ and $\bm{e}_{\rm d}$ in $|e_j \cos \varpi_j|$, $|e_j \sin \varpi_j| \leq 0.10$ ($j = {\rm c}, {\rm d}$)
at the grid-spacing of $0.01$,
and find one minimum of $\chi^2_{\rm c}$ for both $m_{\rm d} = 106 M_{\oplus}$ and $m_{\rm d} = 73 M_{\oplus}$ cases.
Next, for all the sets of ($e_{\rm c},\varpi_{\rm c},  e_{\rm d},\varpi_{\rm d}$) 
in $2\sigma$ ($m_{\rm d} = 106 M_{\oplus}$ case) or $1\sigma$ ($m_{\rm d} = 73 M_{\oplus}$ case) confidence regions around the minimum,
we run integrations varying 
$e_{\rm e} \cos \varpi_{\rm e}$ and $e_{\rm e} \sin \varpi_{\rm e}$ from $-0.1$ to $0.1$ at the grid spacing of $0.01$, 
$m_{\rm c}$ from $0$ to $24 M_{\oplus}$ at the grid spacing of $6 M_{\oplus}$, and
$m_{\rm e}$ from $7$ to $57 M_{\oplus}$ at the grid spacing of $10 M_{\oplus}$ (all of these cover the $1\sigma$ intervals from RV),
to find the set of eight parameters that best fits the TTV of KOI-94d.

\begin{figure}
	\begin{center}
	\includegraphics[width=8.5cm,clip]{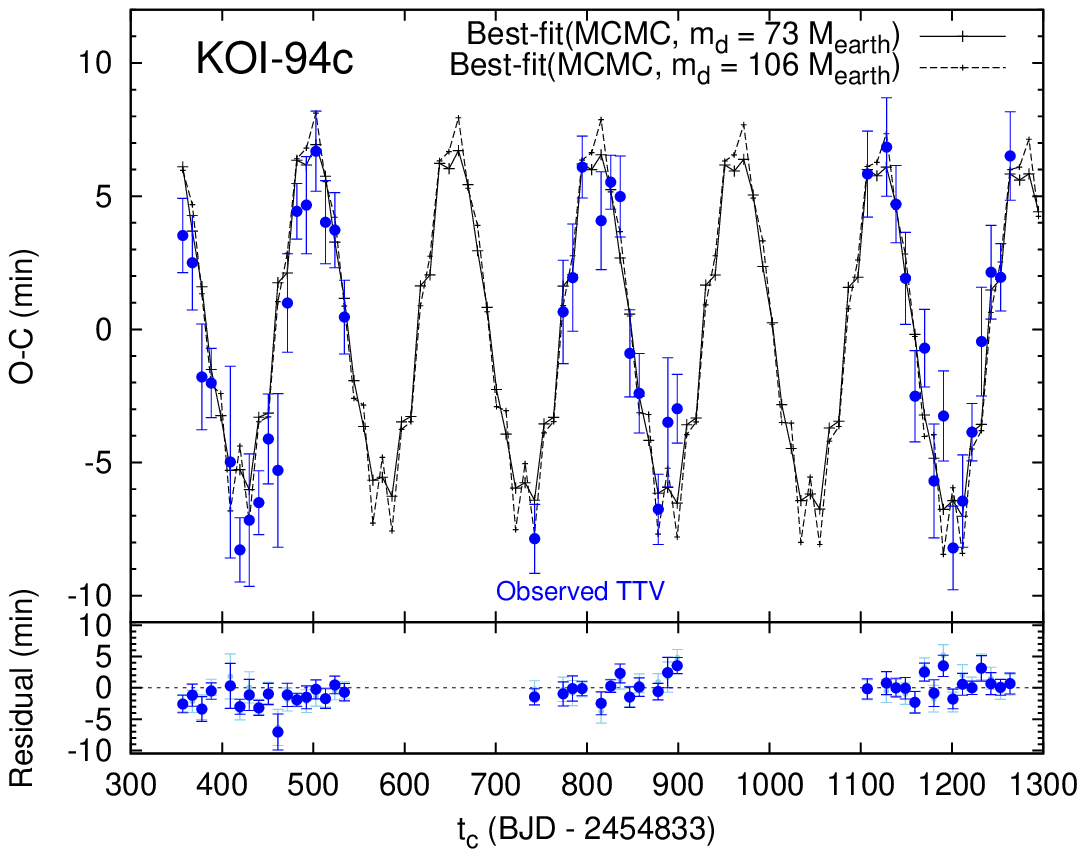}
	\caption{Best-fit simulated TTVs of KOI-94c obtained by the MCMC fit (black crosses connected with lines) with observed data (points with error bars).
	The result for $m_{\rm d} = 73 M_{\oplus}$ is plotted with solid lines (in $O-C$ plot) and blue points (in residual plot),
	and that for $m_{\rm d} = 106 M_{\oplus}$ with dashed lines and sky-blue points.}
	\label{ttv02_rv}
	\end{center}
\end{figure}

\begin{figure}
	\begin{center}
	\includegraphics[width=8.5cm,clip]{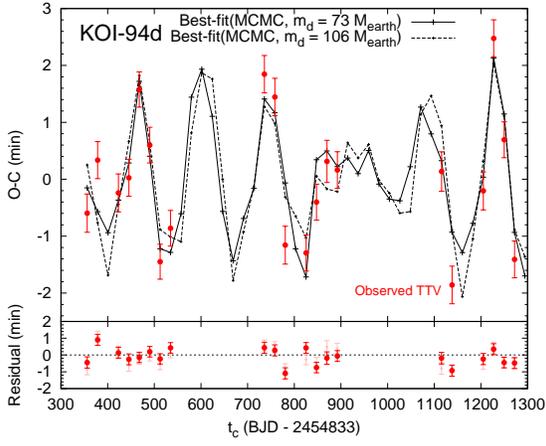}
	\caption{Best-fit simulated TTVs of KOI-94d obtained by the MCMC fit (black crosses connected with lines) with observed data (points with error bars).
	The result for $m_{\rm d} = 73 M_{\oplus}$ is plotted with solid lines (in $O-C$ plot) and red points (in residual plot),
	and that for $m_{\rm d} = 106 M_{\oplus}$ with dashed lines and pink points.}
	\label{ttv01_rv}
	\end{center}
\end{figure}

\noindent{\it MCMC fit to the TTVs of KOI-94c and KOI-94d}.\ ---
Choosing the above set as initial parameters, we then simultaneously fit the TTVs of KOI-94c and KOI-94d using an MCMC algorithm.
In this fit, we use $\chi^2_{\rm c} + \chi^2_{\rm d}$ as the $\chi^2$ statistic.
The resulting best-fit parameters and their $1\sigma$ uncertainties are summarized in Table \ref{constraints_TTV} for the two choices of $m_{\rm d}$
(the second and third columns).
The best-fit simulated TTVs are plotted in Figures \ref{ttv02_rv} and \ref{ttv01_rv} for KOI-94c and KOI-94d, respectively.
\begin{table*}
	\begin{center}
	\caption{Best-fit parameters obtained from TTV analysis.}
	\label{constraints_TTV}
	\begin{tabular}{cccc}
	\tableline \tableline
	Parameter			 		&	Value ($m_{\rm d} = 106 M_{\oplus}$)	&	Value ($m_{\rm d} = 73 M_{\oplus}$)	&	Value (TTV only) \\
	\tableline
	\multicolumn{4}{c}{KOI-94c}\\
	\tableline
	$m_{\rm c}$ ($M_{\oplus}$)		&	$11.8_{-1.5}^{+1.6}$				&	$13.9_{-2.7}^{+2.7}$				&	$9.4_{-2.1}^{+2.4}$\\
	$e_{\rm c} \cos \varpi_{\rm c}$	&	$0.0329_{-0.0055}^{+0.0047}$		&	$0.0092_{-0.0050}^{+0.0264}$		&	$0.0143_{-0.0059}^{+0.0080}$\\
	$e_{\rm c} \sin \varpi_{\rm c}$	&	$-0.0104_{-0.0042}^{+0.0038}$		&	$-0.0031_{-0.0061}^{+0.0067}$		&	$0.0045_{-0.0079}^{+0.0091}$\\
	$\chi^2_{\rm c}$			&	$84$							&	$62$							&	$56$\\
	\tableline
	\multicolumn{4}{c}{KOI-94d}\\
	\tableline
	$m_{\rm d}$ ($M_{\oplus}$)		&	$106$ (fixed)						&	$73$ (fixed)						&	$52.1_{-7.1}^{+6.9}$\\
	$e_{\rm d} \cos \varpi_{\rm d}$&	$0.055_{-0.014}^{+0.011}$			&	$-0.016_{-0.011}^{+0.064}$			&	$-0.022_{-0.011}^{+0.014}$\\
	$e_{\rm d} \sin \varpi_{\rm d}$	&	$0.012_{-0.012}^{+0.011}$			&	$0.009_{-0.018}^{+0.018}$			&	$0.008_{-0.018}^{+0.021}$\\
	$\chi^2_{\rm d}$			&	$66$							&	$48$							&	$43$\\
	\tableline
	\multicolumn{4}{c}{KOI-94e}\\
	\tableline
	$m_{\rm e}$ ($M_{\oplus}$)		&	$15.9_{-2.2}^{+2.4}$				&	$12.9_{-2.3}^{+3.0}$				&	$13.0_{-2.1}^{+2.5}$\\
	$e_{\rm e} \cos \varpi_{\rm e}$	&	$0.067_{-0.019}^{+0.014}$			&	$-0.069_{-0.018}^{+0.120}$			&	$-0.078_{-0.014}^{+0.021}$\\
	$e_{\rm e} \sin \varpi_{\rm e}$	&	$0.042_{-0.017}^{+0.012}$			&	$-0.022_{-0.016}^{+0.032}$			&	$-0.025_{-0.014}^{+0.017}$\\
	\tableline
	$(\chi_{\rm c}^2 + \chi_{\rm d}^2) / {\rm d.o.f}$	&	$150/57$			&	$110/57$							&	$99/56$\\ 
	\tableline 
	\end{tabular}
	\end{center}
\end{table*}

Note that uncertainties of $e_{\rm c} \cos \varpi_{\rm c}$, $e_{\rm d} \cos \varpi_{\rm d}$, and $e_{\rm e} \cos \varpi_{\rm e}$
are relatively large for $m_{\rm d} = 73 M_{\oplus}$ case.
This is because the posterior distributions of these parameters have two peaks, the smaller of which lies close to the best-fit value
for $m_{\rm d} = 106 M_{\oplus}$ case.
Considering this fact, the two results are roughly consistent with each other.
Nevertheless, a total $\chi^2$ in $m_{\rm d} = 73 M_{\oplus}$ case is smaller by $40$ for $57$ d.o.f. than in $m_{\rm d} = 106 M_{\oplus}$ case.
This suggests that the TTV alone favors $m_{\rm d}$ smaller than the RV best-fit value,
as will be confirmed in the next subsection.
\subsection{Solution based only on TTV} \label{sec:Nbody_TTV}
\begin{figure}
	\begin{center}
	\includegraphics[width=8.5cm,clip]{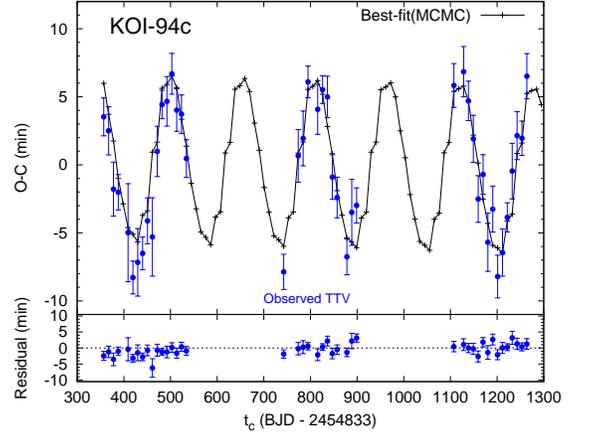}
	\caption{Best-fit simulated TTV of KOI-94c based on the TTV data alone (black crosses connected with lines) with observed data (blue points).}
	\label{ttv_c_ttv}
	\end{center}
\end{figure}

\begin{figure}
	\begin{center}
	\includegraphics[width=8.5cm,clip]{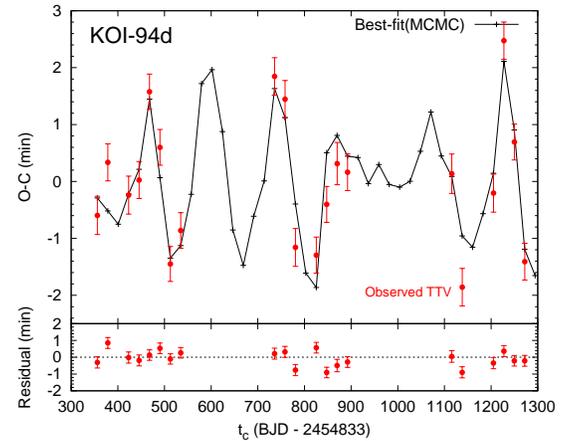}
	\caption{Best-fit simulated TTV of KOI-94d based on the TTV data alone (black crosses connected with lines) with observed data (red points).}
	\label{ttv_d_ttv}
	\end{center}
\end{figure}

In order to obtain a solution independent of RVs, we perform the same MCMC analysis of TTVs of KOI-94c and KOI-94d, 
this time also allowing $m_{\rm d}$ to float.
Since the above analyses suggest that the eccentricities of all the planets are small, 
we choose circular orbits with the best-fit RV masses as an initial parameter set.
The resulting best-fit parameters are summarized in Table \ref{constraints_TTV}, and the corresponding best-fit 
simulated TTVs are shown in Figures \ref{ttv_c_ttv} and \ref{ttv_d_ttv}.
As expected, we find a solution with small eccentricities and with $m_{\rm d}$ smaller than the RV best-fit value.
This solution is similar to that for $m_{\rm d} = 73 M_{\oplus}$ case, except that $m_{\rm d}$ is even smaller.

\subsection{Discussion: comparison with the RV results} \label{sec:ttv_discussion}
While the values of $e_{\rm d}$ and $e_{\rm e}$ obtained in our TTV analysis are consistent with the RV values
in Table \ref{ppar_prior}, the best-fit $e_{\rm c}$ obtained from the TTV is $\sim 1.8\sigma$ smaller than the RV best-fit 
($e_{\rm c} = 0.43 \pm 0.23$).
Considering the marginal detection of KOI-94c's RV and the dynamical stability of the system, 
however, the TTV value seems to be preferred.
In fact, this value is robustly constrained by the clear TTV signal of KOI-94c;
in the grid-search performed in Section \ref{sec:ttv_rv_result}, we searched the region where
$e_{\rm c} \lesssim 0.14$ to fit the TTV of this planet, but the resulting $\chi^2_{\rm c}$ strongly disfavored
large $e_{\rm c}$ regions in both $m_{\rm d} = 106 M_{\oplus}$ and $m_{\rm d} = 73 M_{\oplus}$ cases.

The TTV values of $m_{\rm c}$ and $m_{\rm e}$ are consistent with the RV results, 
but $m_{\rm e}$ is constrained to a rather lower value than the RV best-fit.
Using this value, along with the photometric values of $R_p/R_{\star}$ and $\rho_{\star}$, and spectroscopic value of $M_{\star}$,
the density of KOI-94e is given by $\rho_{\rm e} \sim 0.3\,\mathrm{g\,cm^{-3}}$.
This implies that it is one of the lowest-density planets ever discovered.

The largest discrepancy arises in the value of $m_{\rm d}$, mass of KOI-94d, for which
the TTV best-fit value is smaller than the RV value by $\sim 4\sigma$.
The worse fit in the case of $m_{\rm d} = 106 M_{\oplus}$ is mainly due to the fact that the observed TTV amplitude of KOI-94c
is smaller than expected from this value of $m_{\rm d}$.
As shown in Table \ref{ttv_pair}, the TTV of this planet is completely dominated by the perturbation from KOI-94d, and so
the parameters relevant to this TTV are $m_{\rm d}$, $\bm{e}_{\rm c}$, and $\bm{e}_{\rm d}$.
Table \ref{ttv_pair} also shows that $m_{\rm d} = 106 M_{\oplus}$ leads to the TTV semi-amplitude 
of $\sim 11\,{\rm min}$ for $e_{\rm c} = e_{\rm d} = 0$, which is much larger than the observed TTV amplitude of KOI-94c (see Figure \ref{ttvsignal}).
As a result, the values of $\bm{e}_{\rm c}$ and $\bm{e}_{\rm d}$ are fine-tuned to fit the observed signal,
resulting in strict constraints on these parameters.
The problem is that these values of $\bm{e}_{\rm c}$ and $\bm{e}_{\rm d}$ do not fit the TTV of KOI-94d well.
In the grid-search, we first fit the TTV of KOI-94c alone, and the best-fit gives
$\chi^2_{\rm c} = 66$ for $m_{\rm d} = 106 M_{\oplus}$.
However, this value is largely increased in the simultaneous MCMC fit to the TTVs of KOI-94c and KOI-94d (Table \ref{constraints_TTV}),
which means that these two TTVs cannot be explained with the same set of $\bm{e}_{\rm c}$ and $\bm{e}_{\rm d}$.
On the other hand, this tension does not exist in $m_{\rm d} = 73 M_{\oplus}$ case, in which both of the grid-search and MCMC
return the same values for the best-fit $\chi^2_{\rm c}$.
In fact, the analysis using the analytic formulae of TTVs from two coplanar planets lying near $j:j-1$ resonance \citep{2012ApJ...761..122L}
also supports the above reasoning, suggesting that $m_{\rm d}$ expected from the TTV of KOI-94c is rather small (see Appendix \ref{sec:lithwick}).

For these reasons, it is clear that the TTV favors the solution with 
$m_{\rm d}$ smaller than the RV best-fit value.
It is also true, however, that the RV of KOI-94d is detected with high significance, in contrast to those of the other planets.
Indeed, we calculate the RVs using the best-fit TTV parameters and find that the resulting amplitude is much smaller than
that observed by \citet{2013ApJ...768...14W}.
Since its origin is not yet clear, this discrepancy motivates further investigation of KOI-94
including additional RV/TTV observations.

Finally, we note again that in the above analysis we just fit the {\it TTVs} rather than the {\it transit times} 
of KOI-94c and KOI-94d.
For this reason, our solution corresponds to the planetary orbits whose periods are slightly different from 
the actually observed values.
One may argue that such an approximate method leads to an incorrect solution. 
In order to assure that this is not the case, we fit the transit times of the two planets,
choosing $m$, $a$, $e \cos \varpi$, $e \sin \varpi$, and $\omega + f$ of KOI-94c, KOI-94d, and KOI-94e at time $T_0$
as free parameters (fifteen parameters in total).
We run two MCMC chains starting from
(i) circular orbits with RV best-fit masses and
(ii) a local minimum reached by the Levenberg-Marquardt method \citep{2009ASPC..411..251M}
starting from the TTV best-fit parameters (rightmost column of Table \ref{constraints_TTV}).
We find that the best-fit values of $(m, e, \varpi)$ obtained in these ways show
similar trends as the TTV best-fit in Section \ref{sec:Nbody_TTV}, giving 
comparable  reduced $\chi^2$ values.
Namely, the mass of KOI-94d is much smaller than the RV best-fit and eccentricities of the three planets
are close to zero.

Incidentally, as in the case of the fit to TTVs,
the transit times of KOI-94d are less well reproduced than those of KOI-94c.
In fact, the fit to transit times gives a better $\chi^2_{\rm c}$ than that to TTVs 
because the small linear trend apparent in the lower panel of Figure \ref{ttv_c_ttv} is removed;
on the other hand, $\chi^2_{\rm d}$ values are not so different in both cases.
The difficulty in completely reproducing the orbit of KOI-94d may also indicate the presence of 
the additional perturber discussed in Section \ref{sec:ttv_rv_result}.

\section{Analytic Formulation of the PPE} \label{sec:PPE_form}
In this section, we present a general analytic model of the PPE caused by two planets $1$ and $2$ on circular orbits
(see Appendix \ref{sec:PPE_form_e} for the formulation taking account of $\mathrm{O}(e)$ terms).
In what follows, we use the stellar radius $R_{\star}$ as the unit length because all the observables 
are only related to the lengths normalized to this value.

\subsection{Flux variation due to a PPE}
\begin{figure}
	\begin{center}
	\includegraphics[width=6cm,clip]{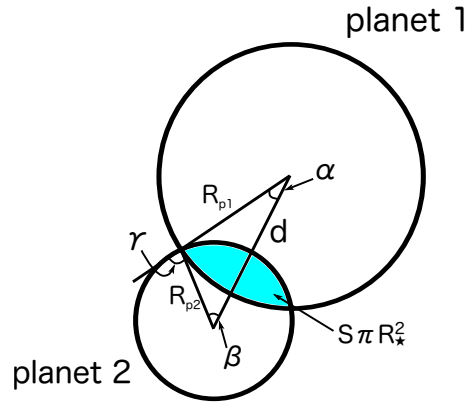}
	\caption{Two overlapping planets $1$ and $2$. The definitions of angles $\alpha$, $\beta$, and $\gamma$ are shown.
	Area of the shaded region corresponds to $S$ in Eq.(\ref{Sab}) times the area of the stellar disk.}
	\label{overlap}
	\end{center}
\end{figure}

A PPE is observed as an increase in the relative flux of a star (or a ``bump" in the light curve) during the double transit of two planets.
Assuming that the two overlapping planets are spherical and neglecting the effect of limb-darkening,
the increase in the relative flux $S$ is given by the area of overlapping region of the two planets 
divided by that of the stellar disk (Figure \ref{overlap}):
\begin{equation}
	S = 
	\begin{cases}
	0 \quad \text{for} \quad  R_{p1} + R_{p2} < d,\\
	\frac{1}{2\pi} R_{p1}^2 (2 \alpha - \sin 2 \alpha) + \frac{1}{2\pi} R_{p2}^2 (2 \beta - \sin 2 \beta)\\
	\ \,\quad \text{for} \quad |R_{p1} - R_{p2}| < d < R_{p1} + R_{p2},\\
	(\min(R_{p1}, R_{p2}))^2 \quad \text{for} \quad d < |R_{p1} - R_{p2}|,
	\end{cases}
	\label{Sab}
\end{equation}
where $d$ is the distance between the centers of the two planets in the plane of the sky,
and the angles $\alpha$ and $\beta$ are defined as
\begin{equation}
	\cos \alpha = \frac{R_{p1}^2 + d^2 - R_{p2}^2}{2 R_{p1} d}, \ 
	\cos \beta = \frac{R_{p2}^2 + d^2 - R_{p1}^2}{2 R_{p2} d}.
	\label{cosab}
\end{equation}
An alternative expression of $S$ for $|R_{p1} - R_{p2}| < d < R_{p1} + R_{p2}$ can be obtained from
the derivative of $S$ with respect to $d$.
Using ${\rm d}\alpha/{\rm d}d$ and ${\rm d}\beta/{\rm d}d$ obtained from Eq.(\ref{cosab})
and $R_{p1} \sin \alpha = R_{p2} \sin \beta$, we obtain
\begin{align}
	\frac{\mathrm{d} S}{\mathrm{d} d} 
	=  - \frac{1}{\pi} \sqrt{ - d^2 + 2 (R_{p1}^2 + R_{p2}^2) - \left( \frac{R_{p1}^2 - R_{p2}^2}{d} \right)^2}
	\label{dsdd}
\end{align}
for $|R_{p1} - R_{p2}| < d < R_{p1} + R_{p2}$.
Eq.(\ref{dsdd}) can be integrated from $R_{p1} + R_{p2}$ to $d$ by changing the variable from
$d$ to
\begin{equation}
	\gamma
	\equiv \alpha + \beta 
	= \pi - \arccos \left(\frac{R_{p1}^2 + R_{p2}^2 - d^2}{2 R_{p1} R_{p2}}\right).
\end{equation}
The result is
\begin{align}
	\notag
	S = \frac{1}{\pi} & 
	\left[ \frac{R_{p1}^2 + R_{p2}^2}{2} \,\gamma - R_{p1} R_{p2} \sin \gamma \right. \\
	&\left.
	- (R_{p1}^2 - R_{p2}^2) \,\arctan \left(\frac{R_{p1} - R_{p2}}{R_{p1} + R_{p2}} \tan \frac{\gamma}{2}\right) \right].
	\label{Sg}
\end{align}
Here, $S$ is given as a function of a single angle $\gamma$.
These two expressions of $S$ show that the shape of a bump due to a PPE is solely
determined by $d$ as a function of time, which will be derived in the following subsection.

The effect of limb-darkening can be included in our model by multiplying $S$ by a factor 
that corresponds to the limb-darkening at the position on the stellar disk over which a PPE occurs.
We adopt the quadratic limb-darkening law 
\begin{equation}
	I(\mu) / I(0) = 1 - u_1 (1 - \mu) - u_2 (1 - \mu)^2,
	\label{qlimb}
\end{equation}
where $\mu = (1 - r^2)^{1/2}$ and $r$ is the radial coordinate on the stellar disk.
For a PPE that occurs totally within the stellar disk, the approximation by \citet{2002ApJ...580L.171M}, 
which is valid for a small planet whose radius is less than about $0.1R_{\star}$,
yields the modified relative increase $S'$ as
\begin{equation}
	S' = S \cdot \frac{I(r_*)}{\int_0^1 dr 2r I(r)}
	= S \cdot \frac{1 - u_1 (1 - \mu_*) - u_2 (1 - \mu_*)^2}{1 - u_1/3 - u_2/6},
	\label{limbfactor}
\end{equation}
where $r_*$ is the distance to the overlapping region.
During the whole PPE, $r_*$ is given in terms of $d$, $\Delta R_{12}^2 \equiv R_{p1}^2 - R_{p2}^2$, and $r_j$ ($j = 1, 2$), 
the radial coordinate of the planet $j$'s center, as 
$r_* = \sqrt{ (r_1^2 + r_2^2)/2 - d^2/4 + (r_2^2 - r_1^2 + \Delta R_{12}^2 /2) \Delta R_{12}^2/2d^2 }$.

\subsection{Distance between the planets during a double transit}
Hereafter, we adopt the Cartesian coordinate system $(X, Y, Z)$ centered on the star,
where the $+Z$-axis is chosen in the direction of our line of sight, and $X$- and $Y$-axes are in the plane of the sky,
forming a right-handed triad.
As stated in the note of Table \ref{ppar_prior},
we align the $+X$ axis with the ascending node of a virtual circular orbit 
whose angular momentum vector is parallel to the stellar spin vector, without loss of generality.
Using $R$ for the three-dimensional distance between the planet and the star, 
the position of a planet is expressed as
\begin{align}
	X &= R \,[ \cos \Omega \cos(\omega + f) - \sin \Omega \sin (\omega + f) \cos i],\\
	Y &= R \,[ \sin \Omega \cos(\omega + f) + \cos \Omega \sin (\omega + f) \cos i],\\
	Z &= R \sin(\omega + f) \sin i.
\end{align}
Suppose that the two planets $1$ and $2$ are on Keplerian orbits 
whose semi-major axes are $a_1$ and $a_2$, respectively.
Neglecting the corrections arising from the non-zero eccentricity, 
the two-dimensional position vectors $\bm{r}_j$ ($j = 1, 2$) of these planets in the plane of the sky 
can be written as
\begin{equation}
	\bm{r}_j = 
	\begin{pmatrix}
	a_j \cos \Omega_j \cos (\omega_j + f_j) - b_j \sin \Omega_j \sin  (\omega_j + f_j)  \\
	a_j \sin \Omega_j \cos  (\omega_j + f_j)+ b_j\cos \Omega_j \sin  (\omega_j + f_j)
	\end{pmatrix}.
	\label{rj}
\end{equation}
If the transits of the two planets are observed, $a_j$, $b_j$, $R_{pj}$, and the periods $P_j$
are obtained as in Section \ref{sec:photometry}.
In this case, the relative motion of the planets is completely described except for the dependence on
the relative nodal angle
defined as
\begin{equation}
	\Omega_{21} = - \Omega_{12} = \Omega_2 - \Omega_1.
\end{equation}

Note that photometric surveys determine the absolute value of $b$, but not its sign.
For a single transiting planet, $b$ is conventionally defined to be positive (or equivalently, $i$ is chosen to be in the range $[0, \pi/2]$),
because the choice of its sign does not affect the transit signals.
For multiple transiting planets, however, 
a different choice of the relative signs of $b$ corresponds to a different orbital configuration.
In this paper, we choose $b_1$ to be positive (i.e. $0 \leq i_1 \leq \pi/2$), 
but allow $b_2$ to be either positive or negative (i.e. $0 \leq i_2 \leq \pi$).
The sign of $b_2$, as well as the relative nodal angle $\Omega_{21}$,
is determined from the observed data of a PPE event.

\begin{figure}
	\begin{center}
	\includegraphics[width=7.5cm,clip]{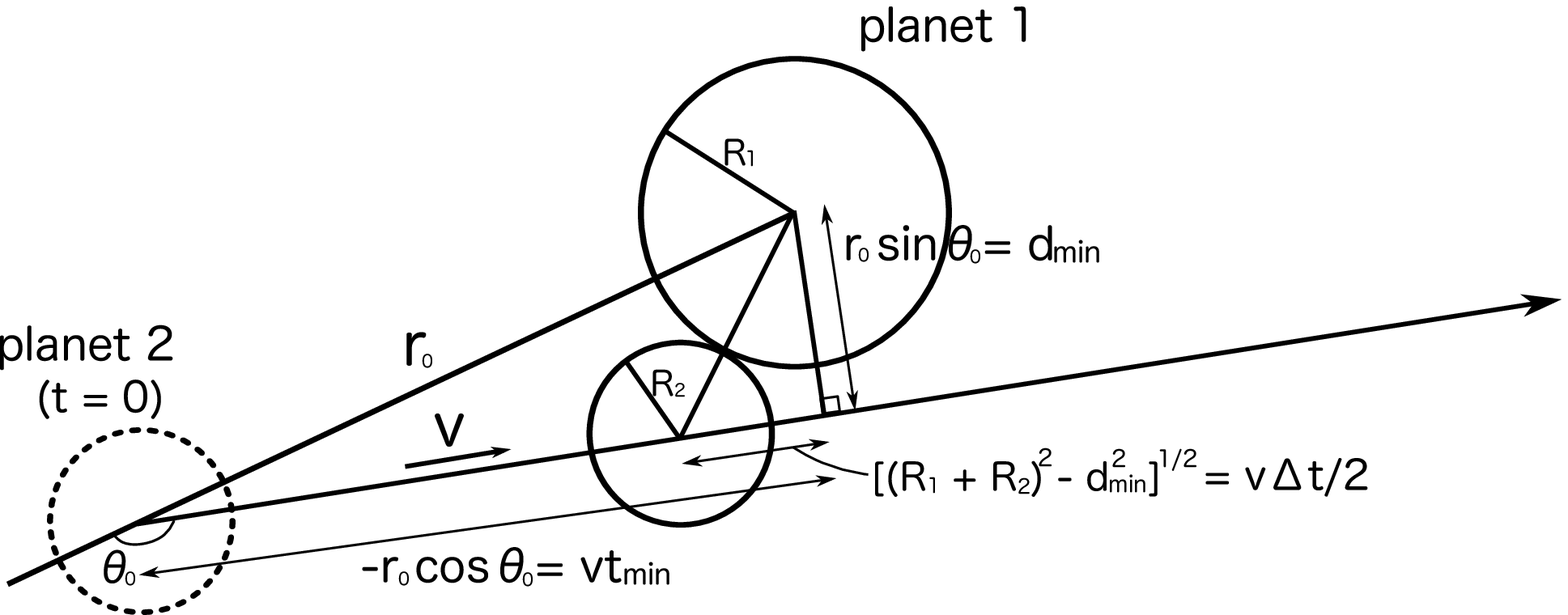}
	\caption{Geometry of the PPE. 
	The relations between $(d_{\mathrm{min}}, t_{\mathrm{min}}, \Delta t)$ and
	$(v, r_0, \theta_0)$ are depicted. 
	Note that the relative position of planet $2$ at $t = 0$ is not its real position at the time,
	but is determined by extrapolating the linear orbit given by Eq.(\ref{r_lin}).
	This geometry shows that a  PPE occurs only when $\cos \theta_0 < 0$,
         in which case $t_{\mathrm{min}}$ given by Eq.(\ref{tmin}) is always positive.}
	\label{geometry}
	\end{center}
\end{figure}

During a double transit by planets $j = 1$ and $2$, their phases at time $t$ are given by
\begin{equation}
	\omega_j + f_j = \frac{\pi}{2} + n_j (t - t_c^{(j)}),
\end{equation}
where $n_j = 2 \pi / P_j$ is the mean motion 
and $t_c^{(j)}$ is the central transit time of planet $j$ in this double transit. 
We then expand Eq.(\ref{rj}) to the first order of $n(t -t_c) \sim \mathcal{O}(a^{-1})$ to obtain
\begin{equation}
	\bm{r}_j = \bm{v}_j (t - t_c^{(j)}) + \bm{r}_0^{(j)},
	\label{r_lin}
\end{equation}
where
\begin{equation}
	\bm{v}_j = - a_j n_j
	\begin{pmatrix}
	\cos \Omega_j \\
	\sin \Omega_j
	\end{pmatrix}, 
	\quad
	\bm{r}_0^{(j)} = -b_j
	\begin{pmatrix}
	\sin \Omega_j \\
	- \cos \Omega_j
	\end{pmatrix}.
	\label{v_r0}
\end{equation}
These give the distance between the two planets $d$ as\footnote{
Strictly speaking, this expression of $d$ contains $\mathcal{O}( (n(t-t_c))^2 )$ terms, 
and so we should use the second-order expansion in Eq.(\ref{r_lin}).
Nevertheless, this only introduces the correction of order $(b/a)^2 \sim a^{-2}$,
which can be safely neglected in our treatment below.}
\begin{equation}
	d^2
	= | \bm{r}_2 - \bm{r}_1 |^2
	= v^2 \left( t + \frac{r_0 \cos \theta_0}{v} \right)^2 + r_0^2 \sin^2 \theta_0,
	\label{d_lin}
\end{equation}
where 
$\bm{v} \equiv \bm{v}_2 - \bm{v}_1$, 
$\bm{r}_0 \equiv \bm{r}_0^{(2)} - \bm{v}_2 t_c^{(2)} - \bm{r}_0^{(1)} + \bm{v}_1 t_c^{(1)}$,
$v \equiv |\bm{v}|$, $r_0 \equiv |\bm{r}_0|$,
and $\theta_0$ is the angle between $\bm{v}$ and $\bm{r}_0$.
Thus, the minimum value of $d$ in this double transit
\begin{equation}
	d_{\mathrm{min}} \equiv r_0 \sin \theta_0,
	\label{dmin}
\end{equation}
is reached at the time
\begin{equation}
	t = t_{\mathrm{min}} \equiv - \frac{r_0 \cos \theta_0}{v}.
	\label{tmin}
\end{equation}
If $d_{\mathrm{min}} < R_{p1} + R_{p2}$,
a PPE occurs during this double transit for a duration of
\begin{equation}
	\Delta t = \frac{2}{v} \sqrt{ (R_{p1} + R_{p2})^2 - d_{\mathrm{min}}^2 }.
	\label{deltat}
\end{equation}
Eqs.(\ref{dmin}) to (\ref{deltat}) can be readily understood by considering the geometry of the PPE
shown in Figure \ref{geometry}.
The explicit expressions of $v$, $r_0$, and $\cos \theta_0$ are 
\begin{align}
	\label{v}
	v^2 &= a_1^2 n_1^2 + a_2^2 n_2^2 -2 a_1 n_1 a_2 n_2 \cos \Omega_{21},\\
	\notag
	r_0^2 &= (a_1 n_1 t_c^{(1)})^2 + (a_2 n_2 t_c^{(2)})^2 + b_1^2 + b_2^2\\
	\notag
	&\quad \quad + 2 ( a_1 n_1 t_c^{(1)} b_2 - a_2 n_2 t_c^{(2)} b_1 ) \sin \Omega_{21}\\
	\label{r_0}
	&\quad \quad - 2 ( a_1 n_1 t_c^{(1)} a_2 n_2 t_c^{(2)} + b_1 b_2 ) \cos \Omega_{21},\\
	\notag
	\cos \theta_0 
	&= \frac{1}{vr_0} \left[ -(a_1 n_1)^2 t_c^{(1)} - (a_2 n_2)^2 t_c^{(2)} \right.\\
	\notag
	&\quad \quad + (a_2 n_2 b_1 - a_1 n_1 b_2) \sin \Omega_{21}\\
	&\quad \quad +\left.  a_1 n_1 a_2 n_2 (t_c^{(1)} + t_c^{(2)}) \cos \Omega_{21}  \right].
	\label{costheta_0}
\end{align}

\subsection{Reconstruction of the mutual inclination $\Omega_{21}$}
The observed shape of a bump is characterized by
its maximum height $S_{\mathrm{max}}$,
central time $t_{\mathrm{min}}$,
and duration $\Delta t$.
If the bump is not saturated, i.e. $d_{\mathrm{min}} > |R_{p1} - R_{p2}|$,
$S_{\mathrm{max}}$ is uniquely translated into $d_{\mathrm{min}}$ via Eq.(\ref{Sab}).
In this case, we can use Eqs.(\ref{dmin}) to (\ref{deltat}), 
the expressions for the three observables of a bump 
($d_{\mathrm{min}}$, $t_{\mathrm{min}}$, $\Delta t$),
to calculate
\begin{align}
	\label{v_obs}
	v &= \frac{2}{\Delta t} \sqrt{(R_{p1} + R_{p2})^2 - d_{\mathrm{min}}^2},\\
	\label{r_0_obs}
	\notag
	r_0^2 &= d_{\mathrm{min}}^2 + (v t_{\mathrm{min}})^2\\
	&= d_{\mathrm{min}}^2 + \left( \frac{2 t_{\mathrm{min}}}{\Delta t} \right)^2 
	\left[ (R_{p1} + R_{p2})^2 - d_{\mathrm{min}}^2 \right],\\
	\tan \theta_0 &= -\frac{d_{\mathrm{min}}}{v t_{\mathrm{min}}}
	= -\frac{\Delta t}{2 t_{\mathrm{min}}} \frac{d_{\mathrm{min}}}{\sqrt{(R_{p1} + R_{p2})^2 - d_{\mathrm{min}}^2}}.
\end{align}
Furthermore, Eqs.(\ref{v}) and (\ref{r_0}), the explicit expressions for $v$ and $r_0$, 
are rewritten as 
\begin{align}
	\label{cosdelta}
	\cos \Omega_{21} &= \frac{a_1^2 n_1^2 + a_2^2 n_2^2 - v^2}{2 a_1 a_2 n_1 n_2},\\
	\label{sindelta}
	\sin \Omega_{21} &= \frac{r_0^2 - x_1^2 - x_2^2 + 2 (\bm{x}_1 \cdot \bm{x}_2) \cos \Omega_{21}}
	{2 | \bm{x}_1 \times \bm{x}_2 |},
\end{align}
where we define
$
	\bm{x}_1 = ( a_1 n_1 t_c^{(1)}, b_1) 
$
and
$
	\bm{x}_2 = ( a_2 n_2 t_c^{(2)}, b_2).
$
In this way, the relative nodal angle $\Omega_{21}$ can be specified explicitly
from ($d_{\mathrm{min}}$, $t_{\mathrm{min}}$, $\Delta t$)
along with the photometrically obtained parameters $(a_j, b_j, n_j, t_c^{(j)}, R_{pj})$,
provided that the sign of $b_2$ is determined.
Although we do not present any general procedure to determine the sign of $b$ here, 
it is possible to do so at least empirically, as described in Section \ref{sec:PPE_app}.
Eq.(\ref{sindelta}) shows that $\sin \Omega_{21}$ is only weakly constrained when $a_1 n_1 / b_1 \simeq a_2 n_2 /b_2$,
in which case the coefficients in front of $\sin \Omega_{21}$ in Eqs.(\ref{r_0}) and (\ref{costheta_0}) are close to zero
because $t_c^{(1)} \simeq t_c^{(2)}$.

In the case of $d_{\mathrm{min}} < |R_{p1} - R_{p2}|$, 
only the upper limit on $d_{\mathrm{min}}$ can be obtained,
and so the entire shape of the bump is required to determine $\Omega_{21}$.

Note that the formulation in this section is also valid even in the presence of non-Keplerian effects.
In such a case, the Keplerian orbital elements in this formulation should be interpreted as
the osculating orbital elements around a certain double transit.

\section{Application to the PPE Observed in the KOI-94 system} \label{sec:PPE_app}
\citet{2012ApJ...759L..36H} fit the whole light curve of the PPE for $u_1$, $u_2$, $t_c^{(1)}$, $t_c^{(3)}$, and $\Omega_{\rm de}$
using an MCMC algorithm, and estimated the relative nodal angle between 
KOI-94d (planet $1$ in Section \ref{sec:PPE_form}) and KOI-94e (planet $2$ in Section \ref{sec:PPE_form}) 
to be $\Omega_{\rm ed} = 1.15 \pm 0.55$ deg.\footnote{
Note that the ``mutual inclination"  $\delta$ defined in \citet{2012ApJ...759L..36H} 
corresponds to $\Omega_{\rm de} = - \Omega_{\rm ed}$ in our definition.}
In their analysis, the light curve is modeled as a sum of two single transit light curves \citep{2009ApJ...690....1O}
and the bump function, which is calculated essentially in the same way as described in Section \ref{sec:PPE_form},
but neglecting the effect of limb-darkening.
In this section, we confirm that this is a unique solution expected from the observed features of the bump,
based on analytic expressions obtained in the previous section.\\
\begin{figure}
	\begin{center}
	\includegraphics[width=7.5cm,clip]{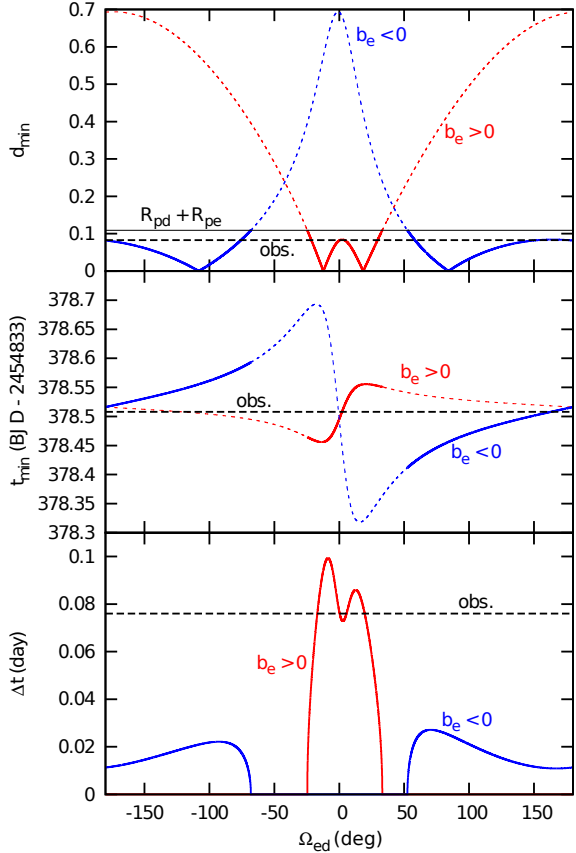}
	\caption{Plots to determine $\Omega_{\rm ed}$ and the sign of $b_{\rm e}$ from the observed bump features.
	(Top) Relation between $d_{\mathrm{min}}$ and $\Omega_{\rm ed}$ based on Eqs.(\ref{dmin}), (\ref{r_0}), and (\ref{costheta_0}). 
	The red and blue lines correspond to $b_{\rm e} > 0$ and $b_{\rm e} < 0$, respectively.
	The solid parts show the region where
	$d_{\mathrm{min}}$ is less than $R_{p\rm{d}} + R_{p\rm{e}}$ (black solid line), i.e.,
	the PPE occurs.
	The dashed black line shows the value of $d_{\mathrm{min}}$ obtained from the 
	best-fit for the observed PPE light curve.
	(Middle) Relation between $t_{\mathrm{min}}$ and $\Omega_{\rm ed}$ based on Eqs.(\ref{tmin}) and (\ref{v}) to (\ref{costheta_0}).
	(Bottom) Relation between $\Delta t$ and $\Omega_{\rm ed}$ based on Eqs.(\ref{deltat}), (\ref{dmin}), and (\ref{v}) to (\ref{costheta_0}). }
	\label{plot_delta}
	\end{center}
\end{figure}
The top panel of Figure \ref{plot_delta} plots $d_{\mathrm{min}}$ as a function of $\Omega_{\rm ed}$ in the double transit during which the PPE was observed,
calculated by Eqs.(\ref{dmin}), (\ref{r_0}), and (\ref{costheta_0}).
In Figure \ref{plot_delta},  we fix $a_j$, $b_j$, $n_j$, and $R_{pj}$ at the values publicized by the Kepler team (Table \ref{ppar_prior}),
and $t_c^{(j)}$ at the best-fit values obtained by \citet{2012ApJ...759L..36H}.
The red and blue lines correspond to the cases of $b_{\rm e} > 0$ and $b_{\rm e} < 0$, respectively,
and the solid parts of the lines show the range of $\Omega_{\rm ed}$ 
for which the PPE occurs, i.e., $d_{\rm min} < R_{p\rm d} + R_{p\rm e}$.
This panel implies that the PPE itself could have occurred 
for a wide range of $\Omega_{\rm ed}$ except for those around $\pm 50$ deg.
The central time $t_{\mathrm{min}}$ and duration $\Delta t$ for a specific value of $\Omega_{\rm ed}$ 
can be obtained from the middle and bottom panels.
These also indicate that a wide variety of bumps could have resulted.

Indeed, these three observables have enough information to reconstruct the value of $\Omega_{\rm ed}$,
in addition to the sign of $b_{\rm e}$.
As for the observed eclipse, the best-fit light curve yields 
$S_{\mathrm{max}} \approx 3.88 \times 10^{-4}$,
$t_{\mathrm{min}} \approx 378.508 \,\mathrm{day}$ $(\mathrm{BJD} - 2454833)$,
and $\Delta t \approx 0.076 \,\mathrm{days}$.
Eq.(\ref{Sab}) shows that the above value of $S_{\mathrm{max}}$ uniquely translates into
$d_{\mathrm{min}} \approx 0.0829$. 
The values of these $d_{\mathrm{min}}$, $t_{\mathrm{min}}$, and $\Delta t$ are plotted in
Figure \ref{plot_delta} in horizontal dashed lines.\footnote{
The analysis that includes the effect of limb-darkening by Eq.(\ref{limbfactor})
returned the same values of $d_{\mathrm{min}}$, $t_{\mathrm{min}}$, and $\Delta t$ 
with a slightly different $\Omega_{\rm ed} \sim 1.21\,\mathrm{deg}$,
in which case the following discussion is also valid.}
For the observed value of $d_{\mathrm{min}}$, Figure \ref{plot_delta} allows eight solutions,
four for each of $b_{\rm e} > 0$ and $b_{\rm e} < 0$.
However, the asymmetry of $t_{\mathrm{min}}$ curve in the middle panel of Figure \ref{plot_delta} shows that
only the solutions around 
$\Omega_{\rm ed} \sim 0$ deg (slightly positive, $b_{\rm e} > 0$) 
or $\Omega_{\rm ed} \sim 180$ deg ($b_{\rm e} < 0$) are possible.
These correspond to the nearly parallel and anti-parallel planetary orbits, respectively.
This degeneracy can be broken with the value of $\Delta t$:
the retrograde (anti-parallel) orbit results in a much shorter bump due to the larger relative velocity between the planets
than the prograde (parallel) case.
The bottom panel of Figure \ref{plot_delta} shows that the observed duration allows only the prograde orbit with $b_{\rm e} > 0$.
In this way, 
$\Omega_{\rm ed} = 1.15$ deg (and $b_{\rm e} > 0$)
proves to be the unique solution that reproduces the observed features of the bump.
In fact, one can show that any set of $(d_{\mathrm{min}}, t_{\mathrm{min}}, \Delta t)$ allows the
unique determination of $\Omega_{\rm ed}$ in the case discussed here.
Mathematically, this comes from the fact 
that the curve $(d_{\mathrm{min}}, t_{\mathrm{min}}, \Delta t)$ parametrized by $\Omega_{\rm ed}$
has no self-intersection.

Combining $\Omega_{\rm ed}$ with the result of the spin-orbit angle measurement,
both $\Omega_{\rm d}$ and $\Omega_{\rm e}$ can be constrained.
Since we have assumed that $b_{\rm d} > 0$ or $0 \leq i_{\rm d} \leq \pi/2$, 
the observed spin-orbit angle $\lambda$ is equal to $\Omega_{\rm d}$ in our definition, 
and so $\Omega_{\rm d} = -6^{+13}_{-11}\,{\rm deg}$.
Thus, $\Omega_{\rm ed} = \Omega_{\rm e} - \Omega_{\rm d} = 1.15 \pm 0.55$ deg \citep{2012ApJ...759L..36H}
gives $\Omega_{\rm e} = -5^{+13}_{-11}\,{\rm deg}$ (Table \ref{ppar_prior}).
Using these two parameters along with the transit parameters in Table \ref{ppar_prior},
we trace the orbits of KOI-94d and KOI-94e for one hundred years,
assuming that their orbits never change over time.
The result of this calculation indicates that the next PPE will occur 
in the double transit around BJD = 2461132.4 (date in UT 2026 April 1/2),
which is the third double transit after the one discussed here.
The same conclusion is obtained even when we vary $\Omega_{\rm ed}$
within its $3\sigma$ interval.
In a real system, however, it is not at all obvious that the next PPE will occur in this double transit, because orbital elements do change over time;
in the next section, we discuss how the mutual gravitational interaction among the planets affects this result.

\section{Implication for the Next PPE: the Effect of Multi-body Interaction} \label{sec:ppnext}
In Section \ref{sec:photometry}, we obtained a set of system parameters by analyzing photometric light curves of the three planets.
Based on this result, we now address the question whether
the PPE will occur in the same double transit as predicted in Section \ref{sec:PPE_app},
even in the presence of the mutual gravitational interaction among the planets.

\subsection{Fixing double-transit parameters} \label{sec:dtparameters}
In order to determine the relative nodal angle crucial in predicting the next PPE,
we refit the transit light curve around BJD = 2454211.5, in which the PPE was observed.
We model the light curve as described in Section \ref{sec:PPE_app} including the effect of limb-darkening
to obtain $R_{p\mathrm{d}}$, $R_{p\mathrm{e}}$, $a_{\rm d}$, $a_{\rm e}$, $b_{\rm d}$, $b_{\rm e}$, $t_c^{\rm (d)}$, $t_c^{\rm (e)}$, and $\Omega_{\rm ed}$.
Here, the same prior as used in fitting the phase-folded light curve 
is assumed to fix the value of $a_{\rm e}$.
We adopt the limb-darkening coefficients obtained from the light curve of KOI-94d,
$u_1 = 0.40$ and $u_2 = 0.14$, which are determined with the best precision 
among the three planets. 
The result is summarized in Table \ref{double_dfix}.\\
\begin{table*}
	\begin{center}
	\caption{Best-fit parameters for the PPE light curve.}
	\label{double_dfix}
	\begin{tabular}{c@{\hspace{3cm}}c}
	\tableline \tableline
	Parameter 					&		 Best-fit value\\
	\tableline
	$R_{p\mathrm{d}} / R_{\star}$				&		$0.06954 \pm 0.00043$\\
	$R_{p\mathrm{e}} / R_{\star}$				&		$0.04123 \pm 0.00070$\\
	$a_{\rm d} / R_{\star}$				&		$26.21 \pm 0.082$\\
	$a_{\rm e} / R_{\star}$				&		$47.44 \pm 0.027$\\
	$b_{\rm d}$						&		$0.2951 \pm 0.0092$\\
	$b_{\rm e}$						&		$0.3693 \pm 0.0095$\\
	$t_c^{\rm (d)}$ (${\rm BJD} - 2454833$)		&		$378.51372 \pm 0.00023$\\
	$t_c^{\rm (e)}$ (${\rm BJD} - 2454833$)		&		$378.51785 \pm 0.00060$\\
	$\Omega_{\rm ed}$ (deg)			&		$1.13 \pm 0.52$\\
	$\chi^2/{\mathrm{d.o.f}}$		&		$746/687$\\
	\tableline 
	\end{tabular}
	\end{center}
\end{table*}
In this fit, $R_{p\mathrm{d}}/R_{\star}$ is different from our revised mean value 
($R_{p\rm{d}}/R_{\star} = 0.07029_{-0.00015}^{+0.00014}$) by $1.7 \sigma$.
We suspect that this difference comes from the systematics introduced in the detrending procedure.
When an artifact or some other astrophysical processes (e.g. star spots) accidentally
increase the relative flux just before (or after) the transit,
the result of detrending is biased towards such features
in setting the baseline of the transit light curve.
If we use a wider range of data points, the effect of such a small feature is averaged out and
does not change the result so significantly.
In contrast, if a narrow region around a transit is used, 
the baseline is somewhat distorted
and the resulting detrended light curve becomes either deeper or shallower.
Such systematics are averaged in the phase-folded light curves, but may be significant 
in an individual transit.
In the case of the double transit light curve analyzed here, the relative flux 
begins to increase just before the ingress, and the depth of the transit is shallower
in the first half of the transit than in the latter, making the light curve slightly asymmetric.
This feature, along with the fact that the revised $R_{p\mathrm{d}}$ gives better $\chi^2$
in all the other transit light curves than $R_{p\mathrm{d}}$ given by the Kepler team,
suggests that the discrepancy in $R_{p\mathrm{d}}$ is caused by such an incidental brightening.
To test this scenario, we repeat the analysis above changing the span of detrending from $\sim 1$ day to $\sim 0.5$ days,
and find that the resulting mean transit parameters are consistent with those above, 
but the depth of the double-transit light curve becomes deeper, consistently with our revised parameters.
\subsection{Evaluation of the multi-body effect}
The occurrence of the PPE in the multi-body context can be assessed in a similar way as in Section \ref{sec:PPE_app};
we compare $R_{p\mathrm{d}} + R_{p\mathrm{e}}$ with $d_{\rm min}$ calculated from Eqs.(\ref{dmin}), (\ref{r_0}), and (\ref{costheta_0}),
but this time the variation of orbital elements must be taken into account.
Specifically, we need to evaluate the variations of the parameters relevant to $d_{\rm min}$, i.e.,
scaled semi-major axis $a/R_{\star}$, mean motion $n$, transit center of the double transit $t_c$,  impact parameter $b$,
and nodal angle $\Omega$ (as long as the eccentricities are small).
In order to give an estimate for these variations, we integrate the orbits of the three planets using the best-fit ($m$, $e$, $\varpi$) 
in the rightmost column of Table \ref{constraints_TTV} up to ${\rm BJD} = 2461132.4$, 
the double transit in which the PPE is expected from the two-body analysis in Section \ref{sec:PPE_app}.
The results are the following:
\begin{enumerate}
\item The oscillation amplitudes of $a_{\rm d}$ and $a_{\rm e}$ are 
less than $5 \times 10^{-5}\,\mathrm{AU}$ ($\sim 0.03\%$) and $4 \times 10^{-4}\,\mathrm{AU}$ ($\sim 0.13\%$), respectively.
Since these are much smaller than the observed uncertainties of $a_{\rm d}/R_{\star}$ and $a_{\rm e}/R_{\star}$ ($\sim 1\%$), 
the multi-body effect can be neglected for these two parameters.
\item Corresponding to the oscillations in semi-major axes above, $n_{\rm d}$ and $n_{\rm e}$ 
also show the modulations whose peak-to-peak amplitudes are $\sim 2\pi (0.01\,{\rm day})^{-1}$ and $\sim 2 \pi (0.1\,{\rm day})^{-1}$,
respectively. These are much larger than the uncertainties in $n_{\rm d}$ and $n_{\rm e}$ that come
from those in $P_{\rm d}$ and $P_{\rm e}$, and so the multi-body effect is important for these parameters.
\item The differences of the periods calculated from the transit centers in the first $\sim 1000$ days (the range in which we analyzed the TTVs)
of integration and from the whole orbit are at most comparable to the observed uncertainties of these parameters in Table \ref{ep}.
Thus the uncertainties of $t_c^{\rm (d)}$ and $t_c^{\rm (e)}$ can be evaluated using those of $P$ and $t_0$ in the table.
Note that the effect of TTV is taken into account in obtaining these errors.
\item Monotonic increase in $i_{\rm d}$ and decrease in $i_{\rm e}$ lead to at most $\sim 30\,\%$ variations of $b_{\rm d}$ and $b_{\rm e}$, 
larger than the observed errors. The multi-body effect is dominant for these parameters.
\item $\Omega_{\rm d}$/$\Omega_{\rm e}$ also monotonically increases/decreases, but only by $< 0.03 \,{\rm deg}$.
This means that the uncertainty in the relative nodal angle $\Omega_{\rm ed}$ is completely dominated by the error in Table \ref{double_dfix},
and the multi-body effect is negligible.
\end{enumerate}

\begin{figure}
	\begin{center}
	\includegraphics[width=7.5cm,clip]{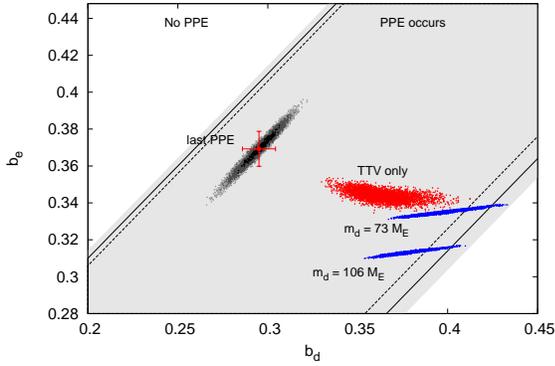}
	\caption{
	Relation between the occurrence of the PPE by KOI-94d and KOI-94e
	and the values of their impact parameters in the double transit around ${\rm BJD} = 2461132.4$.
	If $(b_{\rm d}, b_{\rm e})$ in this double transit is inside the gray-shaded region,
	PPE occurs for all the values of photometrically derived parameters within $1\sigma$ of their best-fit values.
	When we vary them within their $2\sigma$ and $3\sigma$ intervals, 
	the corresponding boundaries become black solid and dashed lines, respectively.
	The values of ($b_{\rm d}$, $b_{\rm e}$) at the last PPE is shown by the red point with error bars,
	and black, dark-gray, and light-gray points around it respectively show their $1\sigma$, $2\sigma$, and $3\sigma$ confidence regions.
	Red and blue points are the distributions of ($b_{\rm d}$, $b_{\rm e}$) in the double transit around ${\rm BJD} = 2461132.4$ 
	for the three different solutions in Table \ref{constraints_TTV}.
	The scattering of each set of points reflects the uncertainties in $(m, e, \varpi)$.}
	\label{ppnext}
	\end{center}
\end{figure}

The above results indicate that the multi-body effect is the most significant for $b_{\rm d}$ and $b_{\rm e}$.
In order to relate the values of these parameters to the occurrence of the PPE during the double transit around ${\rm BJD} = 2461132.4$, 
we use Eq.(\ref{dmin}), (\ref{r_0}), and (\ref{costheta_0}) to calculate the maximum value of $d_{\rm min}$ during this double transit in terms of $b_{\rm d}$ and $b_{\rm e}$, 
varying
(i) $a_{\rm d}$, $a_{\rm e}$, $t_c^{\rm (d)}$, $t_c^{\rm (e)}$, and $\Omega_{\rm ed}$ within $1\sigma$ intervals calculated from the photometric errors, and
(ii) $n_{\rm d}$ and $n_{\rm e}$ by the amplitudes of modulations estimated above.
The region of ($b_{\rm d}$, $b_{\rm e}$) plane in which $d_{\rm min} < R_{p\mathrm{d}} + R_{p\mathrm{e}}$, i.e., the PPE occurs in this double transit,
is shown in Figure \ref{ppnext} with light-gray shade.\footnote{Here, the small uncertainties in $R_{p\mathrm{d}}$ and $R_{p\mathrm{e}}$ are neglected.}
When we vary the parameters in the set (i) within their $2\sigma$ and $3\sigma$ intervals,
the edges of the shaded region can be as narrow as black solid and dashed lines.
In fact, the gray-shaded region is mainly determined by the difference between $b_{\rm d}$ and $b_{\rm e}$, as seen from this figure,
and the edge of this region is found to be most sensitive to the uncertainties in $t_c^{\rm (d)}$ and $t_c^{\rm (e)}$.
The former fact originates from the well-aligned orbital planes of KOI-94d and KOI-94e:
since their orbital planes are nearly parallel in the plane of the sky,
the minimum separation during the double transit in which KOI-94d overtakes KOI-94e
is nearly the same as the difference between $b_{\rm d}$ and $b_{\rm e}$.
However, if their transit times are too far away from each other, such closest encounter may occur out
of the double transit. This explains the latter feature.
The PPE occurrence for different choices of ($m$, $e$, $\varpi$) can be judged 
by evaluating the variation of $b_{\rm d}$ and $b_{\rm e}$ in this plot.
Since the variations of $a$, $n$, $t_c$, and $\Omega_{\rm ed}$ would be of the same order as long as the resulting TTVs are consistent with the observation,
we fix the shaded area in Figure \ref{ppnext} determined by these parameters.
We perform a similar MCMC calculation as in Section \ref{sec:Nbody_TTV} to obtain the distribution of ($b_{\rm d}$, $b_{\rm e}$) in the double transit
at issue; we fit the observed TTVs again using the same $\chi^2$, but this time extend the orbit integration 
up to ${\rm BJD} = 2461132.4$ and record the final values of $b_{\rm d}$ and $b_{\rm e}$ calculated via $b = a \cos i /R_{\star}$.\footnote{
Here we adopt $R_{\star} = (1.37 \pm 0.02) R_{\odot}$ obtained from
photometric $a_{\rm d}/R_{\star}$, $P_{\rm d}$, and spectroscopic $M_{\star}$. 
This value is slightly smaller than $R_{\star} = (1.52 \pm 0.14) R_{\odot}$ obtained by \citet{2013ApJ...768...14W} using the Spectroscopy Made Easy \citep{1996A&AS..118..595V}, 
but this difference is consistent with the conclusion of \citet{2012ApJ...757..161T} that the $\log g$ value based on the Spectroscopy Made Easy is systematically underestimated for stars with
$T_{\rm eff} \gtrsim 6,000\,\mathrm{K}$.}
The resulting distribution of ($b_{\rm d}$, $b_{\rm e}$) is plotted with red points in Figure \ref{ppnext}.
We also repeat the same procedures fixing $m_{\rm d} = 106 M_{\oplus}$ and $m_{\rm d} = 73 M_{\oplus}$,
and the distributions for these cases are plotted with blue points for comparison.
In these calculations, we choose the initial values of $i_{\rm d}$ and $i_{\rm e}$ based on $(b_{\rm d}, b_{\rm e}) = (0.2951, 0.3693)$ in Table \ref{double_dfix},
a red point with error bars in Figure \ref{ppnext},
rather than the mean values obtained from the phase-folded light curves in Table \ref{ppar_fit}.
This is because the mean parameters do not take account of the actual occurrence of the PPE.
The black, dark-gray, and light-gray points around the double-transit value in Figure \ref{ppnext} respectively show
its $1\sigma$, $2\sigma$, and $3\sigma$ confidence regions based on the posterior distribution of the double-transit fit.
Here, the difference between $b_{\rm d}$ and $b_{\rm e}$ is rather sharply constrained by the minimum separation between the planets,
namely, the height of the bump caused by the PPE.
Even considering this uncertainty in the initial $(b_{\rm d}, b_{\rm e})$, as well as the significant variation of impact parameters (or inclinations) due to the multi-body effect, 
$(b_{\rm d}, b_{\rm e})$ around ${\rm BJD} = 2461132.4$ are well inside the region where the PPE occurs,
at least within $1\sigma$ of transit and TTV parameters for all the three solutions.

\begin{figure}
	\begin{center}
	\includegraphics[width=7.5cm,clip]{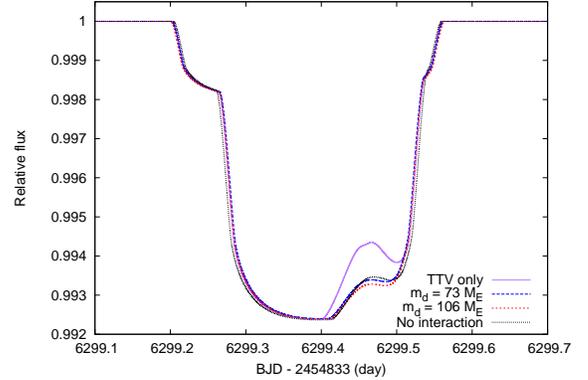}
	\caption{
	The expected light curves of the PPE in the double transit around ${\rm BJD} = 2461132.4$ (date in UT 2026 April 1/2).
	The black dotted curve (No interaction) shows the result for the parameters in Table \ref{double_dfix}.
	The other curves use the median values of $(b_{\rm d}, b_{\rm e})$ for the three distributions in Figure \ref{ppnext},
	with other parameters fixed at the same values as above.}
	\label{ppnext_light}
	\end{center}
\end{figure}

For the best-fit ($m$, $e$, $\varpi$) obtained from the TTV alone, the expected height of the bump is much larger than in the last PPE (Figure \ref{ppnext_light}),
and so the detection of this PPE is highly feasible.
In contrast, for ($m$, $e$, $\varpi$) based on the RV values of $m_{\rm d}$, the bump height is comparable to the last PPE.
This difference may be used to settle the difference of $m_{\rm d}$ values in RV and TTV analyses.
For the RV-based solutions, the blue distributions in Figure \ref{ppnext} indicate that 
the PPE may not even happen when the variation of $b_{\rm d}$ is too large (corresponding to the large $m_{\rm c}$ values),
and/or $\gtrsim 2 \sigma$ deviations of $t_c^{\rm (d)}$ and $t_c^{\rm (e)}$ from the linear ephemerides make the shaded region too narrow
for the PPE to occur (see solid and dashed lines).

We also check the other two double transits before the one discussed above
(around ${\rm  BJD} = 2457982$ and ${\rm  BJD} = 24583612$), 
in case that the variations of $b_{\rm d}$ and $b_{\rm e}$ lead to the PPE which never happens without the interaction
among the planets.
In both of them, we find that the PPE does not happen for any possible values of $b_{\rm d}$ and $b_{\rm e}$ (from $0$ to $1$).
Therefore, we can safely conclude that the next PPE will still occur during the same double transit
as predicted by the two-body calculation, even when we include the mutual gravitational interaction among the planets.

\section{Summary and Discussion} \label{sec:summary}
We have performed an intensive TTV (transit timing variation) analysis
in KOI-94, the first multi-planetary system exhibiting the
PPE \citep[planet-planet eclipse,][]{2012ApJ...759L..36H}.  
Comparison of the resulting system parameters with those
estimated independently from the RV (radial velocity) data
\citep{2013ApJ...768...14W} works as a valuable test to examine the reliability
and limitation of the TTV analysis for other planetary systems for which
the RV data are difficult to obtain. Furthermore, a possible
discrepancy between the two estimates, if any, would be even useful in
exploring additional planets or other interesting implications \citep[e.g.][]{2012Sci...336.1133N}.

Among the four planets reported so far, we considered the TTVs of
KOI-94c, KOI-94d, and KOI-94e; we made sure that the contribution from
the innermost and smallest planet KOI-94b is negligible at the current
level of observational uncertainties.

We numerically integrated the orbits of the three planets that
are directly incorporated in the MCMC search for the best-fit values of
their masses, eccentricities, and longitudes of periastrons; our
best-fit values include $m_{\rm c} = 9.4_{-2.1}^{+2.4} M_{\oplus}$,
$m_{\rm d} = 52.1_{-7.1}^{+6.9} M_{\oplus}$, $m_{\rm e} =
13.0_{-2.1}^{+2.5} M_{\oplus}$, and $e \lesssim 0.1$ for all the three
planets.  Those results are in reasonable agreement with the RV results
\citep{2013ApJ...768...14W}, but we would like to note here a few
possible interesting points.
\begin{enumerate}
\item Although the RV analysis results in a fairly large eccentricity for
KOI-94c ($e_{\rm c} = 0.43 \pm 0.23$), the TTVs indicate a significantly
      smaller value. In fact, the stability analysis of the system 
favors the TTV result.
\item The TTV best-fit value of $m_{\rm d}$
differs from the RV result $m_{\rm d} = 106 \pm 11
M_{\oplus}$ by $\sim 4\sigma$ level.
If the TTV value is correct,
KOI-94d may be inflated, in contrast to the conclusion obtained by
\citet{2013ApJ...768...14W}.
\item The TTV of the outermost planet KOI-94e is not well reproduced in the
      current modeling with the three planets. This might
suggest the presence of additional planets and/or
      minor bodies that have evaded the detection so far.
\end{enumerate}
It is definitely premature to draw any decisive conclusions at this
point. Nevertheless, the above possible discrepancies between the TTV and
RV analyses point to the importance of future follow-up observations of
the KOI-94 system.

In addition, we constructed an analytic model of the PPE.  We derived a practical
approximate formula that explicitly yields the difference
between the longitudes of ascending nodes (mutual inclination in the
plane of the sky) of the two planets in terms of the observed height,
central time, and duration of the brightening caused by the PPE.  We
showed that the PPE light curve observed in the KOI-94 system indeed gives a unique
solution for the mutual inclination.  The effect of the non-zero
eccentricities is taken into account in the formulation described in
Appendix \ref{sec:PPE_form_e}, though it is safely neglected for
the KOI-94 system. Combining the TTV best-fit parameters and our
analytic PPE model, the next PPE in this system is predicted to occur in the
double transit around ${\rm BJD} = 2461132.4$ (date in UT 2026 April 1/2).
The occurrence of the next PPE is robust against the $1\sigma$
uncertainties of the parameters.  Since the predicted height of the bump
is much larger than the last one, the detection of this PPE is highly
feasible.  Indeed, the predicted height of the next PPE sensitively
changes with the value of $m_{\rm d}$. Thus the observation 
may be used to distinguish between the TTV and RV solutions.

\acknowledgments 
This work is based on the photometry of KOI-94 provided by NASA's {\it Kepler} mission,
and the authors express special thanks to the {\it Kepler} team.
K.M. is supported by the Leading Graduate Course for Frontiers of Mathematical Sciences and Physics.
The work by T.H. is supported by Japan Society for Promotion of Science (JSPS) Fellowship for Research (No.\,25-3183).
A.T. acknowledges the support from Grant-in-Aid for Scientific Research by JSPS (No.\,24540257).
M.N. is supported by Grant-in-Aid for Scientific Research on Innovative Areas (No.\,23103005).
Y.S. gratefully acknowledges the supports from the Global Collaborative Research Fund ``A World-wide Investigation of Other Worlds"
grant, the Global Scholars Program of Princeton University, and the Grant-in Aid for Scientific Research by JSPS (No.\,24340035).  
\appendix
\section{Observed transit times of KOI-94${\rm \c}$, KOI-94${\rm \d}$, and KOI-94${\rm \e}$} \label{sec:tc}
In Section \ref{sec:photometry}, we fit each transit of KOI-94c, KOI-94d, and KOI-94e for the time of transit center, using the transit parameters in Table \ref{ppar_fit}.
The resulting transit times of the three planets, as well as their errors, $\chi^2$ values, and deviations from the linear ephemerides in Table \ref{ep}
are shown in Tables \ref{94.02tc} to \ref{94.03tc} in this Appendix.
\begin{table}
	\begin{center}
	\caption{Transit times of KOI-94${\rm \c}$.}
	\label{94.02tc}
	\begin{tabular}{cccccc}
	\tableline \tableline
	Transit number &  $t_c$ ${\rm (BJD-2454833)}$ & $1\sigma_\mathrm{lower}$ & $1\sigma_\mathrm{upper}$ & $\chi^2$/d.o.f & $O-C$ (days)\\
	\tableline
	$	21	 $      &       $	356.90817 	 $      &       $	0.00102 	 $      &       $	0.00092 	 $      &       $	1.10 	 $      &       $	0.00245 	$  \\
	$	22	 $      &       $	367.33115 	 $      &       $	0.00110 	 $      &       $	0.00136 	 $      &       $	1.17 	 $      &       $	0.00174 	$  \\
	$	23	 $      &       $	377.75187 	 $      &       $	0.00142 	 $      &       $	0.00133 	 $      &       $	0.97 	 $      &       $	-0.00124 	$  \\
	$	24	 $      &       $	388.17539 	 $      &       $	0.00094 	 $      &       $	0.00086 	 $      &       $	1.15 	 $      &       $	-0.00140 	$  \\
	$	26	 $      &       $	409.02071 	 $      &       $	0.00231 	 $      &       $	0.00269 	 $      &       $	1.08 	 $      &       $	-0.00346 	$  \\
	$	27	 $      &       $	419.44211 	 $      &       $	0.00082 	 $      &       $	0.00085 	 $      &       $	1.02 	 $      &       $	-0.00575 	$  \\
	$	28	 $      &       $	429.86657 	 $      &       $	0.00197 	 $      &       $	0.00148 	 $      &       $	1.02 	 $      &       $	-0.00497 	$  \\
	$	29	 $      &       $	440.29072 	 $      &       $	0.00080 	 $      &       $	0.00087 	 $      &       $	1.24 	 $      &       $	-0.00452 	$  \\
	$	30	 $      &       $	450.71607 	 $      &       $	0.00120 	 $      &       $	0.00115 	 $      &       $	1.00 	 $      &       $	-0.00286 	$  \\
	$	31	 $      &       $	461.13894 	 $      &       $	0.00231 	 $      &       $	0.00170 	 $      &       $	1.10 	 $      &       $	-0.00368 	$  \\
	$	32	 $      &       $	471.56699 	 $      &       $	0.00124 	 $      &       $	0.00133 	 $      &       $	1.17 	 $      &       $	0.00069 	$  \\
	$	33	 $      &       $	481.99307 	 $      &       $	0.00072 	 $      &       $	0.00073 	 $      &       $	1.05 	 $      &       $	0.00308 	$  \\
	$	34	 $      &       $	492.41692 	 $      &       $	0.00116 	 $      &       $	0.00138 	 $      &       $	1.18 	 $      &       $	0.00324 	$  \\
	$	35	 $      &       $	502.84201 	 $      &       $	0.00107 	 $      &       $	0.00102 	 $      &       $	1.13 	 $      &       $	0.00464 	$  \\
	$	36	 $      &       $	513.26385 	 $      &       $	0.00101 	 $      &       $	0.00115 	 $      &       $	1.14 	 $      &       $	0.00279 	$  \\
	$	37	 $      &       $	523.68733 	 $      &       $	0.00101 	 $      &       $	0.00095 	 $      &       $	1.05 	 $      &       $	0.00259 	$  \\
	$	38	 $      &       $	534.10875 	 $      &       $	0.00100 	 $      &       $	0.00093 	 $      &       $	1.22 	 $      &       $	0.00032 	$  \\
	$	58	 $      &       $	742.57675 	 $      &       $	0.00088 	 $      &       $	0.00091 	 $      &       $	1.11 	 $      &       $	-0.00546 	$  \\
	$	61	 $      &       $	773.85373 	 $      &       $	0.00147 	 $      &       $	0.00122 	 $      &       $	1.17 	 $      &       $	0.00045 	$  \\
	$	62	 $      &       $	784.27831 	 $      &       $	0.00165 	 $      &       $	0.00115 	 $      &       $	1.17 	 $      &       $	0.00135 	$  \\
	$	63	 $      &       $	794.70489 	 $      &       $	0.00076 	 $      &       $	0.00085 	 $      &       $	1.13 	 $      &       $	0.00423 	$  \\
	$	65	 $      &       $	815.55087 	 $      &       $	0.00127 	 $      &       $	0.00128 	 $      &       $	1.10 	 $      &       $	0.00283 	$  \\
	$	66	 $      &       $	825.97556 	 $      &       $	0.00069 	 $      &       $	0.00072 	 $      &       $	0.94 	 $      &       $	0.00384 	$  \\
	$	67	 $      &       $	836.39887 	 $      &       $	0.00097 	 $      &       $	0.00114 	 $      &       $	1.01 	 $      &       $	0.00346 	$  \\
	$	68	 $      &       $	846.81847 	 $      &       $	0.00101 	 $      &       $	0.00126 	 $      &       $	1.13 	 $      &       $	-0.00063 	$  \\
	$	69	 $      &       $	857.24112 	 $      &       $	0.00111 	 $      &       $	0.00096 	 $      &       $	1.08 	 $      &       $	-0.00167 	$  \\
	$	71	 $      &       $	878.08547 	 $      &       $	0.00094 	 $      &       $	0.00089 	 $      &       $	1.09 	 $      &       $	-0.00469 	$  \\
	$	72	 $      &       $	888.51143 	 $      &       $	0.00146 	 $      &       $	0.00191 	 $      &       $	1.16 	 $      &       $	-0.00243 	$  \\
	$	73	 $      &       $	898.93547 	 $      &       $	0.00092 	 $      &       $	0.00087 	 $      &       $	1.07 	 $      &       $	-0.00207 	$  \\
	$	93	 $      &       $	1107.41537 	 $      &       $	0.00097 	 $      &       $	0.00127 	 $      &       $	1.01 	 $      &       $	0.00405 	$  \\
	$	95	 $      &       $	1128.26345 	 $      &       $	0.00112 	 $      &       $	0.00145 	 $      &       $	1.11 	 $      &       $	0.00475 	$  \\
	$	96	 $      &       $	1138.68565 	 $      &       $	0.00111 	 $      &       $	0.00091 	 $      &       $	1.06 	 $      &       $	0.00326 	$  \\
	$	97	 $      &       $	1149.10740 	 $      &       $	0.00118 	 $      &       $	0.00121 	 $      &       $	0.94 	 $      &       $	0.00133 	$  \\
	$	98	 $      &       $	1159.52802 	 $      &       $	0.00122 	 $      &       $	0.00116 	 $      &       $	1.02 	 $      &       $	-0.00174 	$  \\
	$	99	 $      &       $	1169.95296 	 $      &       $	0.00095 	 $      &       $	0.00108 	 $      &       $	1.10 	 $      &       $	-0.00049 	$  \\
	$	100	 $      &       $	1180.37319 	 $      &       $	0.00136 	 $      &       $	0.00161 	 $      &       $	1.00 	 $      &       $	-0.00395 	$  \\
	$	101	 $      &       $	1190.79857 	 $      &       $	0.00119 	 $      &       $	0.00116 	 $      &       $	1.14 	 $      &       $	-0.00226 	$  \\
	$	102	 $      &       $	1201.21882 	 $      &       $	0.00100 	 $      &       $	0.00119 	 $      &       $	1.19 	 $      &       $	-0.00570 	$  \\
	$	103	 $      &       $	1211.64373 	 $      &       $	0.00122 	 $      &       $	0.00119 	 $      &       $	1.06 	 $      &       $	-0.00448 	$  \\
	$	104	 $      &       $	1222.06922 	 $      &       $	0.00076 	 $      &       $	0.00073 	 $      &       $	1.17 	 $      &       $	-0.00268 	$  \\
	$	105	 $      &       $	1232.49526 	 $      &       $	0.00147 	 $      &       $	0.00137 	 $      &       $	1.29 	 $      &       $	-0.00032 	$  \\
	$	106	 $      &       $	1242.92076 	 $      &       $	0.00113 	 $      &       $	0.00132 	 $      &       $	0.96 	 $      &       $	0.00149 	$  \\
	$	107	 $      &       $	1253.34432 	 $      &       $	0.00091 	 $      &       $	0.00085 	 $      &       $	1.17 	 $      &       $	0.00136 	$  \\
	$	108	 $      &       $	1263.77117 	 $      &       $	0.00136 	 $      &       $	0.00095 	 $      &       $	1.17 	 $      &       $	0.00452 	$  \\
	\tableline 
	\end{tabular}
	\end{center}
\end{table}
\begin{table}
	\begin{center}
	\caption{Transit times of KOI-94${\rm \d}$.}
	\label{94.01tc}
	\begin{tabular}{cccccc}
	\tableline \tableline
	Transit number &  $t_c$ ${\rm (BJD-2454833)}$ & $1\sigma_\mathrm{lower}$ & $1\sigma_\mathrm{upper}$ & $\chi^2$/d.o.f & $O-C$ (days)\\
	\tableline
	10		&	$	356.17032 	$	&	$	0.00023 	$	&	$	0.00023 	$	&	$	1.10 	$	&	$	-0.00041 	$	\\
	11		&	$	378.51394$\tablenotemark{a} 		&	$	0.00023 	$	&	$	0.00023 	$	&	$	1.25 	$	&	$	0.00023 	$	\\
	13		&	$	423.19948 	$	&	$	0.00023 	$	&	$	0.00023 	$	&	$	1.17 	$	&	$	-0.00017 	$	\\
	14		&	$	445.54263 	$	&	$	0.00022 	$	&	$	0.00023 	$	&	$	1.14 	$	&	$	0.00002 	$	\\
	15		&	$	467.88668 	$	&	$	0.00021 	$	&	$	0.00022 	$	&	$	1.06 	$	&	$	0.00109 	$	\\
	16		&	$	490.22897 	$	&	$	0.00022 	$	&	$	0.00022 	$	&	$	1.07 	$	&	$	0.00042 	$	\\
	17		&	$	512.57051 	$	&	$	0.00021 	$	&	$	0.00021 	$	&	$	1.17 	$	&	$	-0.00101 	$	\\
	18		&	$	534.91389 	$	&	$	0.00022 	$	&	$	0.00022 	$	&	$	1.08 	$	&	$	-0.00060 	$	\\
	27		&	$	736.00250 	$	&	$	0.00023 	$	&	$	0.00023 	$	&	$	1.00 	$	&	$	0.00128 	$	\\
	28		&	$	758.34519 	$	&	$	0.00023 	$	&	$	0.00023 	$	&	$	1.10 	$	&	$	0.00100 	$	\\	
	29		&	$	780.68635 	$	&	$	0.00023 	$	&	$	0.00023 	$	&	$	1.13 	$	&	$	-0.00080 	$	\\
	31		&	$	825.37220 	$	&	$	0.00022 	$	&	$	0.00022 	$	&	$	1.06 	$	&	$	-0.00090 	$	\\
	32		&	$	847.71579 	$	&	$	0.00022 	$	&	$	0.00022 	$	&	$	1.14 	$	&	$	-0.00028 	$	\\
	33		&	$	870.05926 	$	&	$	0.00026 	$	&	$	0.00026 	$	&	$	1.03 	$	&	$	0.00022 	$	\\
	34		&	$	892.40212 	$	&	$	0.00022 	$	&	$	0.00022 	$	&	$	1.11 	$	&	$	0.00011 	$	\\
	44		&	$	1115.83180 	$	&	$	0.00024 	$	&	$	0.00024 	$	&	$	2.02 	$	&	$	0.00009 	$	\\
	45		&	$	1138.17339 	$	&	$	0.00023 	$	&	$	0.00023 	$	&	$	1.10 	$	&	$	-0.00129 	$	\\
	48		&	$	1205.20344 	$	&	$	0.00023 	$	&	$	0.00023 	$	&	$	1.11 	$	&	$	-0.00014 	$	\\
	49		&	$	1227.54827 	$	&	$	0.00022 	$	&	$	0.00023 	$	&	$	1.01 	$	&	$	0.00172 	$	\\
	50		&	$	1249.89001 	$	&	$	0.00022 	$	&	$	0.00022 	$	&	$	1.08 	$	&	$	0.00048 	$	\\
	51		&	$	1272.23152 	$	&	$	0.00022 	$	&	$	0.00023 	$	&	$	0.94 	$	&	$	-0.00098 	$	\\
	\tableline 
	\end{tabular}
	\tablenotetext{1}{Double transit with KOI-94e: obtained simultaneously with the relative nodal angle and $t_c$ for KOI-94e.}
	\end{center}
\end{table}
\begin{table}
	\begin{center}
	\caption{Transit times of KOI-94${\rm \e}$.}
	\label{94.03tc}
	\begin{tabular}{cccccc}
	\tableline \tableline
	Transit number &  $t_c$ ${\rm (BJD-2454833)}$ & $1\sigma_\mathrm{lower}$ & $1\sigma_\mathrm{upper}$ & $\chi^2$/d.o.f & $O-C$ (days)\\
	\tableline
	$	4	$      &        $	378.51677$\tablenotemark{a} 	     &        $	0.00056 	$      &        $	0.00055 	$      &        $	1.25 	$      &        $	-0.00150 	$ \\
	$	5	$      &        $	432.83744 	$      &        $	0.00062 	$      &        $	0.00063 	$      &        $	1.22 	$      &        $	-0.00068 	$ \\
	$	6	$      &        $	487.15505 	$      &        $	0.00060 	$      &        $	0.00062 	$      &        $	1.11 	$      &        $	-0.00292 	$ \\
	$	11	$      &        $	758.76221 	$      &        $	0.00059 	$      &        $	0.00060 	$      &        $	0.98 	$      &        $	0.00500 	$ \\
	$	12	$      &        $	813.08158 	$      &        $	0.00061 	$      &        $	0.00061 	$      &        $	1.10 	$      &        $	0.00452 	$ \\
	$	18	$      &        $	1138.99524 	$      &        $	0.00060 	$      &        $	0.00062 	$      &        $	1.21 	$      &        $	-0.00091 	$ \\
	$	19	$      &        $	1193.31417 	$      &        $	0.00062 	$      &        $	0.00063 	$      &        $	1.19 	$      &        $	-0.00183 	$ \\
	$	20	$      &        $	1247.63414 	$      &        $	0.00060 	$      &        $	0.00062 	$      &        $	1.24 	$      &        $	-0.00171 	$ \\
	\tableline 
	\end{tabular}
	\tablenotetext{1}{Double transit with KOI-94d: obtained simultaneously with the relative nodal angle and $t_c$ for KOI-94d.}
	\end{center}
\end{table}
\section{Analysis of the TTV of KOI-94${\rm \c}$ using Analytic Formulae} \label{sec:lithwick}
\citet{2012ApJ...761..122L} derived analytic formulae for the TTV signals from 
two coplanar planets near a $j : j-1$ mean motion resonance.
Here we present a brief outline of their formulation and 
report the analysis of the TTVs of KOI-94c and KOI-94d using these formulae.

We let unprimed and primed symbols stand for the quantities associated with inner and outer planets, respectively. 
Then $\delta t \equiv ({\rm observed}\ t_c) - (t_c\ {\rm calculated\ from\ linear\ ephemeris})$ for the inner and outer planets are given by
\begin{equation}
	\delta t = |V| \sin (\lambda^j + \arg V), \quad
	\delta t' = |V'| \sin (\lambda^j + \arg V'),
	\label{ttv_deltat}
\end{equation}
where $\lambda^j$, $V$, and $V'$ are defined as follows.

The longitude of conjunction $\lambda^j$ is defined as
\begin{equation}
	\lambda^j \equiv j \lambda' - (j - 1) \lambda,
\end{equation}
where $\lambda' = 2 \pi (t - T')/P'$ and $\lambda = 2\pi (t - T)/P$.
If we measure angles with respect to the line of sight, $T$ and $T'$ are the times of any particular transits
of the inner and outer planet, respectively. 
Here we choose $T$\,($T'$) to be $t_0$\,($t'_0$) in Table \ref{ep}.
Defining the super-period $P^j$ and the normalized distance to resonance by
\begin{equation}
	P^j \equiv \frac{1}{|j/P' - (j - 1) / P|}
\end{equation}
and
\begin{equation}
	\Delta \equiv \frac{P'}{P} \frac{j - 1}{j} - 1,
\end{equation}
$\lambda^j$ can be written as
\begin{align}
	\notag
	\lambda^j &= - 2 \pi \left( \frac{j-1}{P} - \frac{j}{P'} \right) t + 2 \pi \left( \frac{(j-1)T}{P} - \frac{jT'}{P'} \right)\\
	&= - \frac{\Delta}{|\Delta|}\frac{2 \pi}{P^j} \left( t - \frac{(1 + \Delta) T - T'}{\Delta} \right).
\end{align}
Thus, if $\Delta > 0$, $\lambda^j$ is retrograde with respect to the orbital motion and is prograde for $\Delta < 0$.

The complex TTV amplitudes $V$ and $V'$ are given by
\begin{equation}
	V = P \frac{\mu'}{\pi j^{2/3} (j - 1)^{1/3} \Delta} \left( -f - \frac{3}{2} \frac{Z_{\mathrm{free}}^*}{\Delta} \right)
	\label{ttv_inner}
\end{equation}
and
\begin{equation}
	V' = P' \frac{\mu}{\pi j \Delta} \left( -g + \frac{3}{2} \frac{Z_{\mathrm{free}}^*}{\Delta} \right),
	\label{ttv_outer}
\end{equation}
where $f$ and $g$ are the sums of the Laplace coefficients given by
\begin{equation}
	f = - 1.190 + 2.20 \Delta = -1.032< 0, \quad
	g = 0.4284 - 3.69 \Delta = 0.1637 > 0
\end{equation}
for $j = 2$, $\Delta = 0.07174$, and $\mu$\,($\mu'$) is the mass ratio of the inner (outer) planet to that of the star.
They also introduce $Z_{\mathrm{free}}$ as a linear combination of the free complex eccentricities of the two planets
\begin{equation}
	Z_{\mathrm{free}} = f z_{\mathrm{free}} + g z'_{\mathrm{free}},
\end{equation}
where $z_{\mathrm{free}}$ is defined as the ``free" part of the complex eccentricity
\begin{equation}
	z \equiv e \exp (i \varpi),
\end{equation}
and obtained by subtracting $z_{\mathrm{forced}}$, the forced eccentricity due to the planet's proximity to resonance,
from $z$. 
The forced eccentricities for the inner and outer planets are
\begin{equation}
	\begin{pmatrix}
	z_{\mathrm{forced}}\\
	z'_{\mathrm{forced}}
	\end{pmatrix}
	= - \frac{1}{j \Delta} 
	\begin{pmatrix}
	\mu' f (P / P')^{1/3}\\
	\mu g
	\end{pmatrix}
	\mathrm{e}^{\mathrm{i} \lambda^j}.
\end{equation}
Since $\Delta \gtrsim 0.01$ and $\mu \lesssim 10^{-4}$ typically,
$|z_{\mathrm{forced}}| \lesssim 10^{-2}$, in which case
\begin{equation}
	Z_{\mathrm{free}} \simeq 
	fe \mathrm{e}^{\mathrm{i} \varpi} + ge' \mathrm{e}^{\mathrm{i} \varpi'}
	= (f e \cos \varpi +  g e' \cos \varpi')
	+ \mathrm{i} (f e \sin \varpi + g e' \sin \varpi').
\end{equation}
\citep{2013ApJS..208...22X}.
Note that in either the limit that $|Z_{\mathrm{free}}| \ll |\Delta|$ or $|Z_{\mathrm{free}}| \gg |\Delta|$,
phases of the two planets' TTVs are anti-correlated, as can be seen from the expressions for $V$ and $V'$.
In this case, TTV signals of the two planets provide only three independent quantities, making it impossible
to uniquely determine $|V|$, $|V'|$, $\mathrm{Re}( Z_{\mathrm{free}})$, and $\mathrm{Im}( Z_{\mathrm{free}})$.

Above expressions for $V$ and $V'$ imply that the phases as well as the amplitudes of the two TTV signals
contain important information about their eccentricities.
For ease of discussion, they define
\begin{equation}
	\phi_{\mathrm{ttv}} \equiv \arg \left( V \times \frac{\Delta}{|\Delta|} \right), \quad
	\phi'_{\mathrm{ttv}} \equiv \arg \left( V' \times \frac{\Delta}{|\Delta|} \right).
\end{equation}
With these definitions, $Z_{\mathrm{free}} = 0$ leads to $\phi_{\mathrm{ttv}} = 0\,{\rm deg}$ and $\phi'_{\mathrm{ttv}} = 180\,{\rm deg}$,
independently of the sign of $\Delta$.
In this case, since $\lambda^j$ decreases (increases) with time for $\Delta > 0$ ($\Delta < 0$),
$\delta t$ crosses zero from above (below) whenever $\lambda^j = 0$.
If the observed TTVs have a phase shift with respect to $\phi_{\mathrm{ttv}} = 0\,{\rm deg}$ and $\phi'_{\mathrm{ttv}} = 180\,{\rm deg}$,
this implies that non-zero $Z_{\mathrm{free}}$ exists.
On the other hand, no phase shift does not necessarily mean $Z_{\mathrm{free}} = 0$, for
the phase of $Z_{\mathrm{free}}$ may vanish by chance.
Although it is impossible to judge whether $Z_{\rm free}$ is really zero or not in a single resonant pair with no phase shift,
important conclusions can be obtained by statistical analyses \citep{2013ApJ...772...74W}.

\begin{figure}
	\begin{center}
	\includegraphics[width=7.5cm,clip]{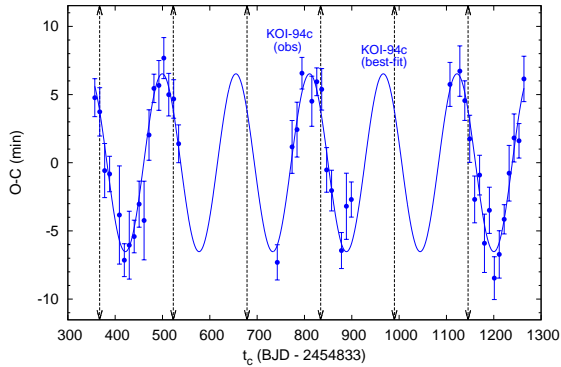}
	\caption{Best-fit theoretical TTV (solid line) for the observed transit times of KOI-94c based on Eq.(\ref{ttv_deltat}) by \citet{2012ApJ...761..122L}.
	Points with error bars are the observed TTVs of KOI-94c calculated with $t_0$ and $P$
	obtained from the fit including TTVs (see Eq.(\ref{t_trans})). Vertical arrows show the times at which $\lambda_j = 0$, i.e., 
	the longitude of conjunction points to the observer. The observed phase of the TTV is slightly shifted from these points,
	suggesting small but nonzero eccentricities.}
	\label{ttv_fit_c}
	\end{center}
\end{figure}

\begin{figure}
	\centering
	\includegraphics[width=7.5cm,clip]{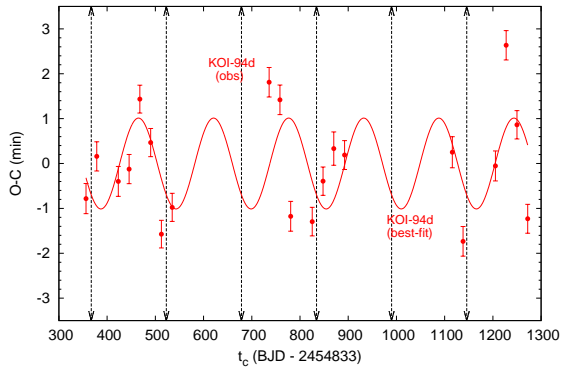}
	\caption{Best-fit theoretical TTV (solid line) for the observed transit times of KOI-94d based on Eq.(\ref{ttv_deltat}) by \citet{2012ApJ...761..122L} (same as Figure \ref{ttv_fit_c}).}
	\label{ttv_fit_d}
\end{figure}

Based on the formulation above, the transit times $t_{\mathrm{trans}}$ for the inner planet are written as
\begin{equation}
	t_{\mathrm{trans}} = t_0 + P i_{\mathrm{trans}} + \mathrm{Re}(V) \sin \lambda^j + \mathrm{Im}(V) \cos \lambda^j,
	\label{t_trans}
\end{equation}
where $i_{\mathrm{trans}} = 0, 1, \cdots$ is the transit number.
For each observed $t_{\mathrm{trans}}$, we calculate $\lambda^j$ using $P$ and $t_0$ obtained by a linear fit (Table \ref{ep}),
and fit for the four parameters $t_0$, $P$, $\mathrm{Re}(V)$, and $\mathrm{Im}(V)$ by a least-square fit.
We also repeat the same procedure for the outer planet, and obtain the results in Table \ref{ttv_lithwick}.
The best-fit theoretical curve in Figure \ref{ttv_fit_c}
shows that the TTV of KOI-94c is well explained only by the effect from KOI-94d,
having the same period as expected from their proximity to 2:1 resonance.
In contrast, the TTV of KOI-94d is poorly explained by the contribution from KOI-94c alone (Figure \ref{ttv_fit_d}).
These results are consistent with our estimates in Table \ref{ttv_pair}.

TTV amplitudes listed in Table \ref{ttv_lithwick} give estimates for the masses of KOI-94d and KOI-94c.
If we assume $Z_{\mathrm{free}} = 0$, i.e., that both of the planets have zero eccentricities,
Eq.(\ref{ttv_inner}) translates the amplitude of KOI-94c's TTV $|V_{\rm c}|$
into the nominal mass $m_{\rm d} = 63 M_{\oplus}.$\footnote{
$|V_{\rm d}|$ corresponds to a comparatively large nominal mass $m_{\rm c} = 36M_{\oplus}$,
but this value includes the contributions both from KOI-94c and KOI-94e.}
However, the accuracy of this estimate is rather limited,
because the slight phase shift in KOI-94c's TTV suggests 
that KOI-94d and/or KOI-94c have small but nonzero eccentricities.
If this nominal mass is actually close to the true one, 
the density of KOI-94d is $\sim 0.3\,\mathrm{g\,cm^{-3}}$, which is comparable
to that of the lowest-density exoplanet ever discovered.

\begin{table}
	\begin{center}
	\caption{Complex TTVs for KOI-94${\rm \c}$ and KOI-94${\rm \d}$.}
	\label{ttv_lithwick}
	\begin{tabular}{ccccccc}
	\tableline \tableline
	$\Delta$		& $|V_{\rm c}|$ (days)	      	& $\phi_{\rm ttv, c}$ (deg) & $|V_{\rm d}|$ (days)		 	&  $\phi_{\rm ttv, d}$ (deg)	& $\chi^2_{\rm c}$/d.o.f 	& $\chi^2_{\rm d}$/d.o.f\\
	\tableline
	 $0.07174$	& $0.0045 \pm 0.0003$ 	& $38 \pm 3$	 		  & $0.00081 \pm 0.00020$	&  $253 \pm 16$ 			& $0.85$				& $8.6$\\
	\tableline 
	\end{tabular}
	\end{center}
\end{table}
\section{$\mathcal{O}(\e)$ Formulation of the PPE} \label{sec:PPE_form_e}
In Section \ref{sec:PPE_form}, we modeled the PPE caused by two planets on circular orbits.
Here we summarize how the $\mathcal{O}(e)$ correction modifies those results.

In the presence of a nonzero eccentricity, the sky-plane distance between the star and planet $r_{\mathrm{sky}}$ is given by
\begin{equation}
	r_{\mathrm{sky}} = \frac{a (1 - e^2)}{1 + e \cos f} \sqrt{1 - \sin^2 ( \omega + f) \sin^2 i}.
\end{equation}
Defining $\Delta \equiv \frac{\pi}{2} - ( \omega + f)$, 
$\Delta$ that minimizes this $r_{\mathrm{sky}}$ satisfies
\begin{equation}
	\Delta = \frac{1}{2} \arcsin \left[
		2 e \cos (\omega + \Delta) \left( \frac{1}{\sin^2 i} - \cos^2 \Delta \right)
		- e \sin (\omega + \Delta) \sin 2 \Delta \right],
\end{equation}
which can be solved to the leading orders of $e$ and $\pi/2 - i$ to give
\begin{equation}
	\Delta \simeq e \cos \omega \left( \frac{\pi}{2} - i \right)^2 \sim e \cos \varpi \left( \frac{a}{R_{\star}} \right)^{-2}.
\end{equation}
Thus, the impact parameter $b$ can be approximated as
\begin{equation}
	b \simeq \frac{a \cos i}{R_{\star}} \cdot \frac{1 - e^2}{1 + e \sin \omega} 
	\simeq \frac{a \cos i}{R_{\star}} \,(1 - e \sin \omega)
\end{equation}
\citep{2010arXiv1001.2010W}.
This alters the expression (\ref{rj}) as
\begin{equation}
	\bm{r}_j 
	\simeq (1 - e_j \cos f_j)
	\begin{pmatrix}
	(a_j/R_{\star}) \cos \Omega_j \cos (\omega_j + f_j)  - b_j (1 + e_j \sin \omega_j) \sin \Omega_j \sin (\omega_j + f_j)\\
	(a_j/R_{\star}) \sin \Omega_j \cos (\omega_j + f_j) + b_j (1 + e_j \sin \omega_j) \cos \Omega_j \sin (\omega_j + f_j)
	\end{pmatrix}.
	\label{r_planetj_e}
\end{equation}
In addition, the expansion of $\omega + f$ around $t_c$ is modified as
\begin{equation}
	\omega + f  \simeq \frac{\pi}{2} + n (1 + 2 e \sin \omega) (t - t_c).
\end{equation}
Using these expressions, $\bm{r}_j$ ($j = 1, 2$) can be expanded as
\begin{equation}
	\bm{r}_j = \bm{v}_j \left[ (1 + e_j \sin\omega_j) (t - t_c^{(j)}) \right]
	+ \bm{r}_0^{(j)},
\end{equation}
where $\bm{v}_j$ and $\bm{r}_0^{(j)}$ are the same as defined in Eq.(\ref{v_r0}).
Accordingly, the expression for $d$ with $\mathcal{O}(e)$ terms included 
is obtained by replacing $b_j$ and $n_j$ 
in the circular case with $b_j (1 - e_j \sin \omega_j)$ and $n_j (1 + e_j \sin \omega_j)$
, respectively.
\bibliography{/Users/kmasuda/work/ApJ_format/reference_masuda.bib}
\end{document}